\documentclass[twoside]{article}
\usepackage{amsfonts,amssymb,amsbsy,textcomp,marvosym,amsmath,caption,threeparttable,amsthm,subfigure}
\usepackage{eurosym,mathrsfs,fancyhdr,CJK,multicol,graphics,indentfirst,color,bm,upgreek,booktabs,graphicx}

\usepackage{multirow}
\usepackage{url}
\def\footnoterule{\kern 1mm \hrule width 10cm \kern 2mm}

\newcommand{\tabincell}[2]{\begin{tabular}{@{}#1@{}}#2\end{tabular}}

\allowdisplaybreaks
\catcode`@=11
\def\title#1{\vspace{3mm}\begin{flushleft}\vglue-.1cm\Large\bf\boldmath\protect\baselineskip=18pt plus.2pt minus.1pt #1
\end{flushleft}\vspace{1mm} }
\def\author#1{\begin{flushleft}\normalsize #1\end{flushleft}\vspace*{-4pt} \vspace{3mm}}
\def\address#1#2{\begin{flushleft}\vglue-.35cm${}^{#1}$\small\it #2\vglue-.35cm\end{flushleft}\vspace{-2mm}\par}

\catcode`@=11
\def\section{\@startsection{section}{1}{\z@}%
 {-3ex \@plus -.3ex \@minus -.2ex}%
 {2.2ex \@plus.2ex}%
{\normalfont\normalsize\protect\baselineskip=14.5pt plus.2pt minus.2pt\bfseries}}
\def\subsection{\@startsection{subsection}{2}{\z@}%
 {-3ex\@plus -.2ex \@minus -.2ex}%
 {2ex \@plus.2ex}%
{\normalfont\normalsize\protect\baselineskip=12.5pt plus.2pt minus.2pt\bfseries}}
\def\subsubsection{\@startsection{subsubsection}{3}{\z@}%
 {-2.2ex\@plus -.21ex \@minus -.2ex}%
 {1.4ex \@plus.2ex}
{\normalfont\normalsize\protect\baselineskip=12pt plus.2pt minus.2pt\sl}}

\begin{document}
\begin{CJK*}{GBK}{song}
\thispagestyle{empty}
\vspace*{-13mm}
\noindent {\small Journal of computer science and technology: JOURNAL OF COMPUTER SCIENCE AND TECHNOLOGY}
\vspace*{2mm}

\title{A Survey of Non-Volatile Main Memory Technologies: State-of-the-Arts, Practices, and Future Directions }

\author{Hai-kun Liu$^{1,2}$, Member, CCF/IEEE/ACM,  Di Chen$^{1,2}$, Hai Jin$^{1,2}$, Fellow, CCF/IEEE, Xiao-fei Liao$^{1,2}$, Member, CCF/IEEE/ACM, Binsheng He$^{3}$, Member, IEEE/ACM, Kan Hu$^{1,2}$ and Yu Zhang$^{1,2}$, Member, CCF/IEEE/ACM}
\address{1}{National Engineering Research Center for Big Data Technology and System, Services Computing Technology and System Laboratory, Cluster and Grid Computing Laboratory, Huazhong University of Science and Technology, Wuhan, 430074, China}
\address{2}{School of Computing Science and Technology, Huazhong University of Science and Technology, Wuhan, 430074, China}
\address{3}{School of Computing, National University of Singapore, 117418, Singapore}
%
%
\noindent E-mail: \{hkliu, chendi, hjin, xfliao\}@hust.edu.cn, hebs@comp.nus.edu.sg, \{hukan, zhyu\}@hust.edu.cn, \\[-1mm]

\let\thefootnote\relax\footnotetext{{}\\[-4mm]\indent\ Regular Paper}

\noindent {\small\bf Abstract} \quad  {\small {\emph{Non-Volatile Main Memories} (NVMMs)  have recently emerged as promising technologies for future memory systems. Generally, NVMMs have many desirable properties such as high density, byte-addressability, non-volatility, low cost, and energy efficiency, at the expense of high write latency, high write power consumption and limited write endurance. NVMMs have become a competitive alternative of \emph{Dynamic Random Access Memory} (DRAM), and will fundamentally change the landscape of memory systems. They bring many research opportunities as well as challenges on system architectural designs, memory management in \textit{operating systems} (OSes), and programming models for hybrid memory systems. In this article, we first revisit the landscape of emerging NVMM technologies, and then survey the state-of-the-art studies of NVMM technologies. We classify those studies with a taxonomy according to different dimensions such as memory architectures, data persistence, performance improvement, energy saving, and wear leveling. Second, to demonstrate the best practices in building NVMM systems, we introduce our recent work of hybrid memory system designs from the dimensions of architectures, systems, and applications. At last, we present our vision of future research directions of NVMMs and shed some light on design challenges and opportunities.}}

\vspace*{3mm}


\vspace*{4mm}

\end{CJK*}
\baselineskip=18pt plus.2pt minus.2pt
\parskip=0pt plus.2pt minus0.2pt
\begin{multicols}{2}

\section{Introduction}\label{sec:introduction}

In-memory computing is becoming increasingly popular for data-intensive applications in the big data era. 
The memory subsystem has an ever-increasing impact on the functionality and performance of modern
computing systems. Traditional big memory systems~\cite{Ousterhout:2015,In-Memory} using DRAM are facing severe scalability challenges in terms of power and density~\cite{DRAMPOWER}. Although DRAM scaling is continued from 28nm in 2013 to 10+nm in 2016~\cite{101,2,3}, the scaling has slowed down and become more and more difficult. Moreover, recent studies~\cite{5,6, 57, 7056524, 6911963} have showed that DRAM-based main memory account for about 30\%-40\% of the total energy consumption of a physical server.

Emerging \textit{Non-Volatile Main Memory} (NVMM) technologies, such as \emph{Phase Change Memory} (PCM), \emph{Spin-Transfer Torque RAM} (STT-RAM), and \textit{3D X-Point} \cite{83} generally offer much higher memory density, much lower cost-per-bit and standby power consumption than DRAM. The advent of NVMM technologies has potential to bridge the gap between slow persistent storage (i.e., disk and SSD) and DRAM, and will fundamentally change the landscape of memory and storage systems. 

\begin{table*}[t]
	\footnotesize
	\caption{Different Features of NVMM Technologies}
	\label{tab1}
	\tabcolsep 20pt 
	\begin{tabular}{|c|c|c|c|c|}
		\hline
		\tabincell{c}{\textbf{Memory} \\ \textbf{Technology}} &  \tabincell{c}{\textbf{Read Latency} \\(ns) }&  \tabincell{c}{\textbf{Write Latency} \\(ns)} &  \tabincell{c}{ \textbf{Write Endurance}\\ (times)} & \tabincell{c}{\textbf{Standby} \\ \textbf{Power}}\\ \hline
		\textbf{ Flash SSD}  & 25,000           & 200,000           & $10^{5}$                  & zero          \\ \hline
		\textbf{DRAM}       & 80                & 80                 & \textgreater$10^{16}$    & Fresh power   \\ \hline
		\textbf{PCM}        & 50-80             & 150-1000           & $10^8$                   & zero           \\ \hline
		\textbf{STT-RAM }   & 6                 & 13                 & $10^{15}$                & zero           \\ \hline
		\textbf{ReRAM}      & 10                & 50                 & $10^{11}$                & zero           \\  \hline
       \tabincell{c}{\textbf{Intel Optane}\\ \textbf{DCPMM}}   & \tabincell{c}{169 (Sequential),\\ 305 (Random)}    & \tabincell{c}{90}   & $10^{8}$                &   zero  \\  \hline
        \hline
	\end{tabular}
\end{table*}

\begin{table*}[t]
\footnotesize
\caption{A Classification of State-of-the-art Studies about NVMM Technologies}
\label{tab2}
\tabcolsep 20pt 
\begin{tabular}{|l|l|}
       \hline
        \multicolumn{2}{|c|}{\textbf{Memory Architectural Studies}}\\
         \hline \hline
        Simulators and Emulators & NVMain~\cite{NVMain},  ZSim~\cite{Sanchez:2013}, HSCC~\cite{HSCC}, PMEP~\cite{38}, HME~\cite{HME}, Quartz~\cite{volos2015quartz}, LEEF~\cite{LeeF} \\
        \hline
       \tabincell{c}{ Hybrid Memory \\ Architectures} &  \tabincell{l}{Horizontal memory architectures~\cite{8,yoon2012row,zhang2009exploring,7,park2011power,CLOCK-DWF,86,7586040,MALRU-DATE,MALRU-TCAD} \\ Hierarchical memory architectures \cite{Qureshi:2009,mladenov2012efficient,loh2011efficiently,meza2012enabling,Liu:2017,TACO2019,ACCESS2020} }  \\
        \hline
        \multicolumn{2}{|c|}{\textbf{OS-level Hybrid Memory Management}}\\
        \hline \hline
        \tabincell{c}{ Persistent Memory \\ Management} & \tabincell{l}{Working Memory:~\cite{8,yoon2012row,zhang2009exploring,7,park2011power,Qureshi:2009,mladenov2012efficient,loh2011efficiently,32,Exploiting};\\ Persistent Memory File System: PMFS~\cite{38},BPFS~\cite{condit2009better}, SCMFS~\cite{5}, SIMFS~\cite{SIMFS}, \\ Contour~\cite{Contour}, Dapper~\cite{Dapper}, NOVA \cite{39}, Orion~\cite{Orion}, ZoFS~\cite{ZoFS}; \\
        Persistent Objects: Mnemosyne~\cite{9}, CDDSs~\cite{16}, NV-Heap~\cite{21}, NV-Duet~\cite{34}, \\NVL-C~\cite{74}, Pangolin~\cite{Pangolin}, TimeStone~\cite{TimeStone}, Pisces~\cite{Pisces}, Espresso~\cite{Espresso} }\\
            \hline
         \tabincell{c}{ Performance Improvement \\ and Energy Saving} & \tabincell{l}{Page Migration: \cite{7,8,CLOCK-DWF,86,93,99,ICCD2019,Exploiting,SCIS2019};\\ Buffering NVMM Writes:~\cite{6,53,87,park2011power};\\ NVMM Energy Saving:~\cite{18,72,73,70,88};\\ DRAM Energy Saving:~\cite{54,55,56,condit2009better, 80} } \\
         \hline
         \tabincell{c}{Write Endurance\\ Improvement} & \tabincell{l}{Write Reduction:~\cite{7,8,57,Qureshi:2009,17,18,36,70,95,82}\\ Wear-Leveling:~\cite{8,58,59,92} }\\
         \hline
         \hline
         \multicolumn{2}{|c|}{\textbf{NVMM Programming Models and Applications}}\\
        \hline
        \tabincell{c}{Programming Models\\ and APIs} & \tabincell{l}{Persistent Objects: Mnemosyne~\cite{9}, CDDSs~\cite{16}, NV-Heap~\cite{21}, NV-Duet~\cite{34}, \\NVL-C~\cite{74}, Pangolin~\cite{Pangolin}, TimeStone~\cite{TimeStone}, Pisces~\cite{Pisces}, Espresso~\cite{Espresso}} \\
        \hline
        \tabincell{c}{Applications using NVMMs} & \tabincell{l}{Key-Value Stores~\cite{HMcached,NVHT,HiLSM,facebook,PapyrusKV,sigmod2019,VLDB-index,Dash-VLDB,mahapatra2019don,ni-ssrctr-20-01,FlatStore}, \\ Graph Computing~\cite{NGraph,GraphNVM,10.14778/3389133.3389145},  Machine Learning~\cite{8715178,Pinatubo,ReRAM-PIM} } \\
        \hline   
\end{tabular}
\end{table*}

Table~\ref{tab1} shows different memory features of Flash SSD, DRAM, PCM, STT-RAM,  ReRAM, and Intel Optane \textit{DC Persistent Memory Modules} (DCPMM)~\cite{Intel-Optane-DIMM} including read/write latencies, write endurance,  and standby power consumption~\cite{6,9,16}. Despite various advantages in density and energy consumption, NVMM exhibits about $6\sim30\times$ higher write latency and about $5 \sim10\times$ higher write power consumption than DRAM. Moreover, the write endurance of NVMM is very limited (about $10^8$ times) while DRAM is able to endure about $10^{16}$ time write operations~\cite{7,8}.  These disadvantages make it hard to be a direct substitute for DRAM. A more practical way of using NVMM is hybrid memory architectures, composed of both DRAM and NVMM~\cite{8,Qureshi:2009}.

In order to fully exploit the advantages of both DRAM and NVMMs in hybrid memory systems, there are many open research problems such as performance improvement, energy saving, cost reduction, wear leveling, and data persistence. To address those problems, there have been many studies on the design of memory hierarchy~\cite{8,Qureshi:2009,yoon2012row,Liu:2017}, memory management~\cite{condit2009better,38,Agarwal:2017}, and memory allocation schemes~\cite{NV-Heaps:2011,Volos:2011,Chakrabarti:2014}. Those research efforts lead to innovations in hybrid memory architecture, operation system (OS), and programming models. Although academic community and industry have proposed a substantial amount of work on integrating the emerging NVMMs in the memory hierarchy, there still remain many challenges to be addressed. 

On the other hand, previous studies on NVMM technologies are mostly based on simulated/emulated NVMM devices. The promised performance of NVMM devices may have various deviations compared to real non-volatile DIMMs. Recently, the announced Intel Optane DCPMM~\cite{Intel-Optane-DIMM} has finally made NVMM DIMMs commercially available. The real Intel Optane DCPMM behaves significantly differently from the promised features that are expected by previous studies. For example,  Intel Optane DCPMMs show 2$\sim$3$\times$ higher read latency than DRAM, while its write latency is even lower than that of DRAM~\cite{fast20}, as shown in Table~\ref{tab1}. The maximal read and write bandwidths for a single Optane DCPMM DIMM are 6.6GB/s and 2.3GB/s, respectively, while the gap between read and write bandwidth of DRAM is much smaller (1.3X). Moreover, the read/write performance are non-monotonic with increasing number of parallel threads in the system~\cite{fast20}. In their experiment, the peak performance is achieved between one and four threads and then tails off. Because of these key features of Optane DCPMM DIMMs,  previous studies on persistent memory systems should be revisited and re-optimized to adapt to the real NVMM DIMMs.

\textbf{Contributions}. In this article, we first revisit the state-of-the-art works on hybrid memory architectures, OS-level hybrid memory management, and hybrid memory programming models. Table 2 shows a classification of state-of-the-art studies about NVMM technologies. We classify these works in a taxonomy according to different dimensions including memory architectures, PM management, performance improvement, energy saving, wear leveling, programming models, and applications. We also discuss their similarities and differences to highlight the design challenges and opportunities. Second, to demonstrate the best practices in building NVMM systems, we present our efforts of hybrid memory system designs from the dimensions of architectures, systems, and applications. At last, we present our vision of future research directions of using NVMMs in real application scenarios, and shed some light on design challenges and opportunities in the research field.  

Although there are other surveys about NVMMs, this survey offers the unique perspective of NVMM and gives more recent review of this field given the rapid development of NVMM. In \cite{ACM-CS}, the authors introduce architectural designs of PCM techniques to address the problems of limited write endurance, potential long latency, high energy writes, power dissipation, and some concerns for memory privacy. In \cite{10.1145/2893186}, the authors present a comprehensive survey and review of PCM device related computer architectures and software. Some other interesting surveys such as \cite{10.1145/3131848} focuses on architecturally integrating four NVM technologies (PCM, MRAM, FeRAM, and ReRAM) into the existing storage hierarchy, or focuses on the software optimizations~\cite{7120149} of using NVMMs for storage and main memory systems.  \emph{Our survey is different from those surveys in three-folds.} First, the articles \cite{ACM-CS,10.1145/2893186} both put a focus on the PCM system designs from the perspective of computer architecture. In contrast, our paper mainly focuses on system works of using hybrid memories from the dimensions of memory hierarchy, system software, and applications. Second, our paper contains more reviews of newly-published journal/conference papers. Particularly, we have provided more studies on the new announced Intel Optane DCPMM device. Third, we introduce more our recent experiences of hybrid memory systems to shed some light on design challenges and opportunities of future hybrid memory systems.

The rest of this paper is organized as follows. Section~\ref{sec:architecture} describes the existing hybrid memory architectures composed of DRAM and NVMMs. Section~\ref{sec:PM} presents the challenges and current solutions of data persistence guarantees in NVMMs.  Section\ref{sec:energy} describes state-of-the-art works on performance optimization and energy saving in hybrid memory systems. Section~\ref{sec:endurance} introduces studies of NVMM write endurance. Section~\ref{sec:practices} presents our efforts and practices of NVMM techniques. In Section~\ref{sec:vision}, we discuss the future research directions of NVMMs, and conclude in Section~\ref{sec:conclusion}.

\section{Hybrid Memory Architectures}\label{sec:architecture}

There have been a lot of studies on hybrid memory architectures. Generally, there are mainly two kinds of hybrid memory architectures, i.e., horizontal and hierarchical~\cite{Liu:2017}, as shown in Figure 1.

\subsection{Horizontal Hybrid Memory Architectures} 
A number of DRAM/NVMM hybrid memory systems~\cite{8,zhang2009exploring,7} manage DRAM and NVMM in a flat (single) memory address space by OSes~\cite{zhang2009exploring,park2011power}, and use both of them as main memory. To improve data access performance, those hybrid memory systems need to overcome the drawbacks of NVMM by migrating frequently accessed (hot) NVMM pages to DRAM, as shown in Figure 1(a). Memory access monitoring mechanisms need to be developed to guide the page migration. 

\begin{center}
	\includegraphics[width=\linewidth]{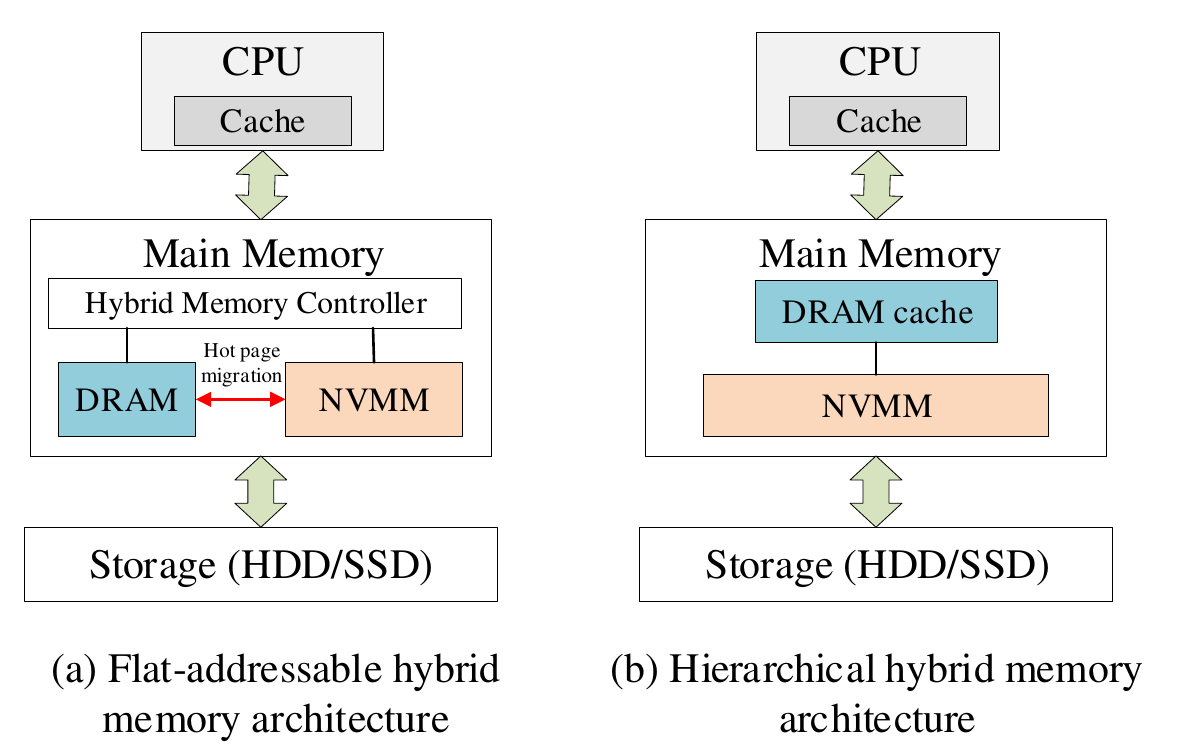}\\
	\vspace{2mm}
	\parbox[c]{8.3cm}{\footnotesize{Fig.1.~} Horizontal and hierarchical hybrid memory architectures}
	\label{fig1}
\end{center}

\emph{Memory access monitoring.} Zhang et al.~\cite{zhang2009exploring} use a multi-queue algorithm to classify the hotness of pages, and place hot pages and cold pages in DRAM and NVMM, respectively. Park et al.~\cite{park2011power} also advocate a horizontal hybrid memory architectures to manage DRAM and NVMM.  Moreover, they propose three optimization strategies to reduce energy consumption of hybrid memory system. They monitor memory data in a very fine granularity of a DRAM row, and periodically check the access counter of each DRAM row. According to counters, the data is written back to NVMM in order to reduce the energy consumption of DRAM refreshing. The data is not cached to DRAM from the NVMM until it is accessed again. The dirty data is kept in DRAM as long as possible to reduce the overhead of data swapping between DRAM and NVMM as well as the costly writes to NVMM. 

\emph{Page migration.}  There have been a number of page migration algorithms proposed for different optimization goals.  Soyoon et al.~\cite{CLOCK-DWF} deem that the frequency of NVMM writes is  more important than data access recency in identifying hot pages, and propose a page replacement algorithm called \textit{CLOCK with Dirty bits and Write Frequency} (CLOCK-DWF). For each NVMM write operation, CLOCK-DWF needs to first fetch the corresponding page to DRAM and then performs the write in DRAM. This approach may cause many unnecessary page migrations, and thus introduce more energy consumption and write-back operations to NVMM. Rezaet al.~\cite{86} take both memory writes and reads into account to migrate the hot pages that are beneficial for performance and energy saving, and use two \textit{Least Recently Used} (LRU) queues to choose victim pages in DRAM and NVM individually. Yoon et al.~\cite{yoon2012row} conduct page migrations based on row buffer locality, where pages with low row buffer hit rates are migrated to DRAM while pages with high row buffer hit rates are still kept in NVMM. Yang et al.~\cite{Utility}  propose a utility model to guide page migrations based on an utility definition on many factors such as page hotness, memory level parallelism and row buffer locality. Khouzani et al.~\cite{7586040} consider memory layout of programs and memory level parallelism to migrate pages in a hybrid memory system. 

\emph{Architectural limitations.} There are several challenges to manage NVMM and DRAM in a horizontal hybrid memory architecture. 

First, page-level memory monitoring is costly. On the one hand, as today's commodity x86 systems do not support memory access monitoring at the granularity of pages, hardware-supported page migration schemes require significant hardware modification to monitor memory access statistics~\cite{8,CLOCK-DWF,7}. On the other hand, memory access monitoring at the OS layer usually cause significant performance overhead. Many OSes maintain an ``accessed'' bit in the \textit{page table entry} (PTE) for each page to identify whether this page is accessed. However, this bit can not truly reflect the recency and frequency of page accesses. Thus, some software-based approaches would disable \textit{Translation Lookaside Buffer} (TLB)~\cite{Gandhi:2014} to track each memory reference. Such page access monitoring mechanisms usually cause significant performance overhead and even offset the benefit of page migration in hybrid memory systems. 

Second, page migration is also costly.  One time of page migration may induce many times of page read/write operations (costly). As a page may only contain a small fraction of hot data, migration at the page granularity is relatively costly due to a waste of memory bandwidth and DRAM capacity. 

Third, the hot page detection mechanism may take a long period of time to allow page to become hot, and thus degrades the gain of page migration. Moreover, the hot page prediction may be not accurate for some irregular memory access patterns, causing unnecessary page migrations.


\subsection{Hierarchical Hybrid Memory Architectures} 
A number of studies propose to organize DRAM and NVMM through a hierarchical cache/memory architecture~\cite{Qureshi:2009,mladenov2012efficient,loh2011efficiently}. They use DRAM as a cache of NVMM, as shown in Figure 1(b). The DRAM cache is invisible to operating systems and applications, and are managed completely by hardware. 

Qureshi~\cite{Qureshi:2009} et al. propose a hierarchical hybrid memory system composed  of a large size of PCM and a small size of DRAM. The DRAM cache contains most recently accessed data to reduce most expensive NVMM accesses, while the large-capacity NVMM holds most of required data during application execution to avoid high-latency disk I/O operations. Similarly, Mladenov~\cite{mladenov2012efficient} et al. design a hybrid memory system with a small-capacity DRAM cache and a large-capacity NVMM, and manage them based on the spatial locality of application data. The DRAM is managed as an on-demand cache and replaced through a LRU algorithm. Loh et al.~\cite{loh2011efficiently} manage DRAM in a granularity of cache lines to improve the efficiency of DRAM cache, and use a group-connected manner to map NVMM data to the DRAM cache. They put the metadata (tag) and data in the same bank row, so that the data can be quickly accessed for cache hits, and reduces the performance overhead of tag querying.

In this memory architecture, as the DRAM is organized as N-way set-associative cache, additional hardware is required to manage the DRAM cache. For example, a SRAM storage is  needed to store the metadata (i.e, tag) of data blocks in the DRAM cache, and hardware looking-up circuit is required to find the requested data in the DRAM cache. Thus, to access data in the DRAM cache, two memory references are required, one for accessing the metadata and the other for the actual data. To accelerate metadata access speed, Qureshi et al.~\cite{Qureshi:2009} use a high-speed SRAM to store the metadata. Meza et al.\cite{meza2012enabling} reduce hardware cost for tag store by placing metadata alongside data blocks in the same DRAM row. They also propose to use a on-chip metadata buffer to cache frequently accessed metadata in a small-size SRAM.

\emph{Architectural limitations.} Although the hierarchical hybrid memory architecture usually deliver much better performance compared with only accessing data in NVMM solely, it may cause significant performance degradation when running workloads with poor locality~\cite{31}. The reason is that most hardware-managed hierarchical DRAM/NVMM systems leverage an on-demand based data fetching policy for simplicity, and thus the DRAM cache is in the critical data path of memory hierarchy. If a data block does not hit in the DRAM cache, it has to be fetched from NVMM to DRAM regardless of the page hotness. This cache filling strategy may cause frequent data swapping between DRAM and NVMM (similar to the cache thrashing problem). On the other hand, hardware-managed cache architecture can not fully utilize the DRAM capacity. Since the DRAM cache is designed to be set-associative, each NVMM data block is mapped to a fixed set. When a set is full, it must evict a data block before fetching a new NVMM data block into the DRAM, even though other cache sets are empty. 

\subsection{Architectures of Intel Optane DCPMM}
The recently announced Intel Optane DCPMM supports both horizontal and hierarchical hybrid memory architectures when it is used combining with DRAM.  There are currently two operating modes for Optane DCPMM DIMMs: \textit{Memory Mode} and \textit{Application Direct Mode} (Persistent)~\cite{fast20}. Each of these modes has its own advantages for specific use cases. 

\textbf{Memory Mode.} In this mode, the DCPMM acts as a large capacity of main memory. The \textit{operating system} (OS) recognizes DCPMM as traditional DRAM and the persistence feature of DCPMM is disabled. If traditional DRAM is used combining with DCPMM, it is hidden from the OS and acts as a caching layer for DCPMM. Thus, the DCPMM and DRAM are actually organized in a hierarchical hybrid memory architecture. The primary benefit of the memory mode is to provide superior memory capacity to be used on memory bus lanes. This mode strongly emphasizes building large storage capacity environments around the memory space \emph{without modifying the upper-level systems and applications}. Recommended use cases would be to expand the main memory capacity for better infrastructure scaling, such as parallel computing platforms for big data applications (mapreduce, graph computing).

\textbf{Application Direct Mode.} In this mode, DCPMM offers all persistence features to the OS and applications. OS exposes both DRAM and DCPMM to the applications as main memory and persistent storage, respectively. The traditional DRAM mixed with DCPMM still acts as standard DRAM for applications, while the DCPMM is also assigned to the memory bus for faster memory access. The DCPMM is used as one of two types of namespaces: \textit{direct access} (DAX) and \textit{block storage}. The former namespace offers byte-addressable persistent storage directly accessed by applications via special APIs. Thus, the DCPMM and DRAM are logically organized in a horizontal hybrid memory architecture in this mode. The latter namespace presents DCPMM to applications as a block storage
device, similar to an SSD, but can be accessed via a faster memory bus.  The Application Direct Mode strongly emphasizes the advantage of latency reduction and bandwidth improvement up to 2.7x faster than NVMe. Recommended use cases would be for large in-memory databases which are subjected to the demand of data persistence. 

There is also a mixed memory mode combining the Memory Mode and the Application Direct Mode. A portion of the capacity of the DCPMM is used for the Memory Mode operations, and the remaining capacity of the DCPMM is used for the Application Direct Mode operations. This mixed memory mode provides a more flexible approach to manage the hybrid memory system for different application scenarios. 

\subsection{Summary} 

The above two kinds of hybrid memory architectures have their own pros and cons for different scenarios. Generally, the hierarchical architecture is more suitable for applications with good data locality, while the flat-addressable architecture is more applicable for latency-insensitive or large-footprint applications. There is not a conclusion on which architecture is better than another one. Actually, both hierarchical and flat-addressable hybrid memory architectures are all supported by the Intel Optane DCPMM~\cite{Intel-Optane-DIMM}. One limitation of current DCPMM is that, the system needs to restart after a reconfiguration on the mode of DCPMM.  It could be interesting and flexible for many applications if a re-configurable hybrid memory system can dynamically fit different scenarios in a timely and efficient manner. This can be an interesting future direction.

\section{Persistent Memory Management}\label{sec:PM}

Data persistence is an important design aspect of NVMM. In the following, we first present the technical challenges of persistent memory (PM) management, and then introduce the state-of-the-art works on PM management, including  the usage of PM, PM access modes,  fault tolerance mechanisms, and persistent objects. 


\subsection{Technical Challenges}
In hybrid memory systems, NVMM can act as main memory when running applications and serve as persistent storage when applications are completed. The byte-addressability and non-volatility features of NVMM eliminate the distinction of memory and external storage. However, the data in NVMM should be reorganized and relocated when the data need to be persisted in the NVMM.


\begin{center}
    \includegraphics[width=0.85\linewidth]{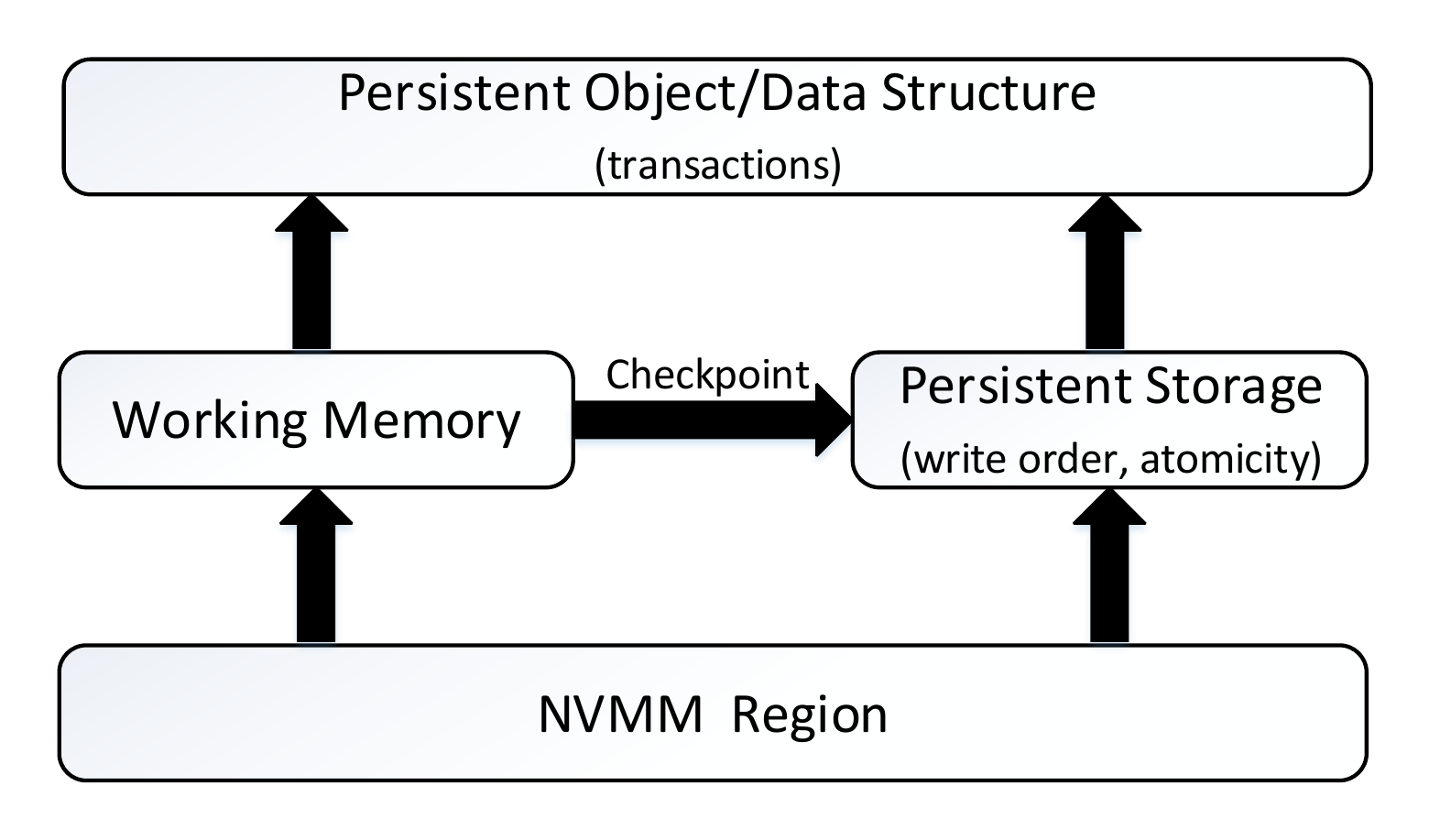}\\
    \vspace{2mm}
    \parbox[c]{8.3cm}{\footnotesize{Fig.2.~} Data Persistence in NVMM}
\end{center}


Figure 2 shows management operations of \textit{persistent memory} (PM). The NVMM region is a physical PM device. The NVMM region can be used as working memory like DRAM, and can also work as persistent storage like disk. When the program is completed, the data in the working memory should be flushed into the persistent storage. Besides, to guarantee high reliability, a checkpointing mechanism is widely exploited to recover system from power failure or system crashes. Gao et. al.~\cite{94} have developed a novel method to leverage NVMM for real-time checkpointing in hybrid memory systems. 


There are several challenges to manage PM efficiently. First, persistent storage is  widely managed in the form of file systems. As the byte-addressable NVMMs offer much better random access performance than traditional block devices, the performance bottleneck of PM-based file systems have shifted from the hardware to the system software stack. It is essential to shorten the data path in the software stack. Second, since many CPUs use write-back cache to achieve high performance for write operations. The last level cache (LLC) may change the order of data written back to PM. In case of a power failure or system crash, it may cause a data inconsistency problem. Thus, to guarantee data consistency in PM, the order of write operations and a write atomicity model are required to guarantee data consistency in PM. Third,  persistent objects and data structures are more promising for PM programming compared to PM based file systems. Because they eliminate the complex data structures in file systems, including i-nodes, metadata, and data. However, these persistent objects and data structures still face the challenges of guaranteeing data consistency.

In the following, we will review the works that have attempted to address those challenges.

\subsection{Working Memory}
A number of studies~\cite{7,81,84} use NVMM just as a replacement of DRAM, without concerning about the non-volatility property of NVMM. In this use case, both DRAM and NVM are allocated and reclaimed in pages. Application data is written back to external storage when programs complete. 


Due to the performance gap between NVMM and DRAM, memory allocation should take the difference features of DRAM/NVMM into account. Park et al. propose to place data in hybrid memory system by exploiting application virtual memory layout~\cite{32}. Both the DRAM region and the NVM region are managed by the buddy system separately. Upon a page fault, the page allocator selects a type of pages for allocation based on the segment in which they are placed. Pages in heap and stack segments with intensive write operations are allocated in DRAM. Pages in other segments are allocated in NVMM, including read-only text segment and initialized data segment. Similarly, Wei et al.~\cite{Exploiting} also exploit application semantics to direct data placement in hybrid memory systems. However, they determine the placement of heap objects based on object read/write ratios. The above memory allocation policies are implemented in OSes and transparent to programmers. The page placement is also too coarse-grained to some extent since programmers often allocate small-size objects rather than pages.

\subsection{Persistent Memory File System}
Some work manages NVMM with traditional file systems to transparently support legacy applications. The file system managed NVMM region is called \emph{Persistent Memory File System} (PMFS)~\cite{38}. In PMFS, applications can access the data in PM via \emph{read}/\emph{write} interfaces as traditional disk-based file systems. The CPU can also directly access PM via \emph{load}/\emph{store} instructions based on \emph{Direct Access} (DAX), which is implemented by the \emph{mmap} interface, as  shown in Figure 3.

Although PM is able to significantly improve application performance compared to  persistent storage, the direct access to byte-addressable PM still faces challenges of data consistency. As an update to a complex data structure usually contains multiple write operations on NVMM, a power failure or a system crash may incur data inconsistency problems if only a portion of critical data is being written. For example, there are two write operations to insert an item to a hash table in PM: one to write the data and the other to write the metadata. If the metadata is persisted before the data itself and a power failure occurs, the data and its metadata become inconsistent. 

Current file systems or databases use atomic updates to tackle this problem, where the correlated write operations are grouped and are performed in a transaction manner, namely transaction updating. Also, in each transaction, multiple writes usually should be constrained in order.


\begin{center}
	\includegraphics[width=\linewidth]{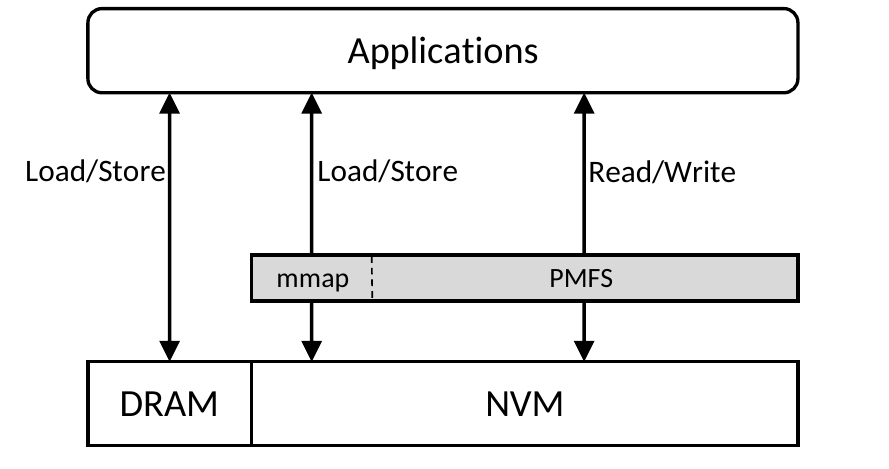}\\
	\vspace{2mm}
	\parbox[c]{8.3cm}{\footnotesize{Fig.3.~} PM access mode in hybrid memory systems}
\end{center}

\subsubsection{Write Order Guarantee}

For block-based file systems, the write order to persistent storage are guaranteed through software due to the huge performance gap between main memory and disk. The I/O operations are buffered sequentially in the DRAM and flushed to persistent storage synchronously. However, in hybrid memory systems, cache lines may be written back to NVMM in an order different from the order issued by CPUs. To guarantee the order of data writes to PM, there are generally three kinds of approaches in the following. 



\textbf{Hardware Primitives.}
PM file systems and programming frameworks can ensure the write order by explicitly evicting cache lines to NVMM. Modern CPUs provide \emph{clflush}  and \emph{mfence} instructions to achieve this goal. The \emph{clflush} instruction is used to evict a cache line to main memory explicitly.  The \emph{mfence} instruction is used to guarantee the order of all \emph{load} and \emph{store} operations before it. There have been various index access methods that optimize the usage of those instructions such as NV-tree~\cite{7271036,7271036} and Bztree~\cite{10.1145/3164135.3164147}.

Nevertheless, \emph{clflush} invalidates a cache line in all cache levels and leads to performance degradation. Moreover, \emph{clflush} only flush a cache line to memory controller, and does not guarantee the data is actually written in NVMM. To tackle these problems, Intel has developed two new enhancement instructions, i.e., \emph{clwb} and \emph{PCOMMIT}~\cite{35}. \emph{clwb} writes back a cache line to memory controllers without invalidating it in the cache, and \emph{PCOMMIT} ensures the data is finally written to the NVMM chips.

\textbf{Write-through Cache.}
Some previous studies adopt the write-through cache to guarantee data persistence in PM~\cite{9,22}. Write operations can bypass CPU caches via instructions like \emph{movntq}. It writes dirty data directly to memory rather than cache, offering a simple way to guarantee the write order to NVMM, without using the complicated barrier and costly flush operations. However, this strategy leads to significant performance degradation because write operations are manipulated in a stream manner. Mnemosyne ~\cite{9} provides both the hardware primitives and write through policies to guarantee the order of writes.

\textbf{Persistent Cache.}
Kiln~\cite{36} utilizes a non-volatile cache as the last level cache (LLC) to guarantee data persistence at the cache level. In the non-volatile LLC, updates can be completed in-place. Kiln tracks the dirty lines that need to be updated but still retain them in the non-volatile cache. As most of the updated writes have existed in the non-volatile LLC, Kiln improved the system performance by reducing most writes to NVMM. Besides, the updated data is kept even when a system failure occur.

\subsubsection{Atomic Updating}
Atomicity implies each update should be  done in a ``all-or-nothing'' manner. It can avoid a data structure being partially updated upon power failures or system crashes. It is always implemented by a transaction operation.
Modern processors can provide 8-byte atomic updates to DRAM or NVMM. An update to a simple variable up to 8 bytes can be done in-place. For more complex data structures, atomic updating operations become more complicated. There are three technologies to guarantee complex atomic operations, such as journaling, shadow updating, and logging structure.


\textbf{Journaling.}
Journaling is commonly used in databases and file systems to guarantee atomic updating. All updates in a transaction are recorded to a journal file before the real object is updated. Thus, journaling always writes the same data twice, one to the journal file and the other for the actual data. To diminish the performance overhead due to duplicate writes, most systems only record metadata in journal files.
For example, Ext4-DAX \cite{37} supports direct access to NVMM and uses the journaling mechanism to achieve metadata atomicity.

\textbf{Shadow Paging.}
Shadow paging is a \textit{copy-on-write} (COW) mechanism for tree-based file systems and databases. Each write operation triggers a memory copy.
In the context of file systems, the COW operation needs to transfer from the root to the leaf in a cascade way. This cascade updating is performance costly. In BPFS ~\cite{condit2009better}, Jeremy et al. propose a short-circuit shadow paging mechanism for atomic updating. Data updates are token in-place, including in-place updating and in-place appending. Fig. 4 shows an example of in-place appending. The appended data is written to the end of the file in-place, and then the file size is updated in-place. In case of a system crash before the updating of file size, the appended data is invalid. In PMFS ~\cite{38}, DRAM pages are allocated and reclaimed by virtual memory manager and NVMM pages are managed by PMFS. Atomic updating is achieved through three approaches: in-place, logging and COW. In-place updates are used for 8-byte metadata atomic writes and metadata updates at 64-byte cache line size. Logging updates are used for more complicated metadata updates. The COW mechanism is used to update file data.


\begin{center}
	\includegraphics[trim={0.1cm 0.1cm 0.1cm 0.1cm},clip, width=\linewidth]{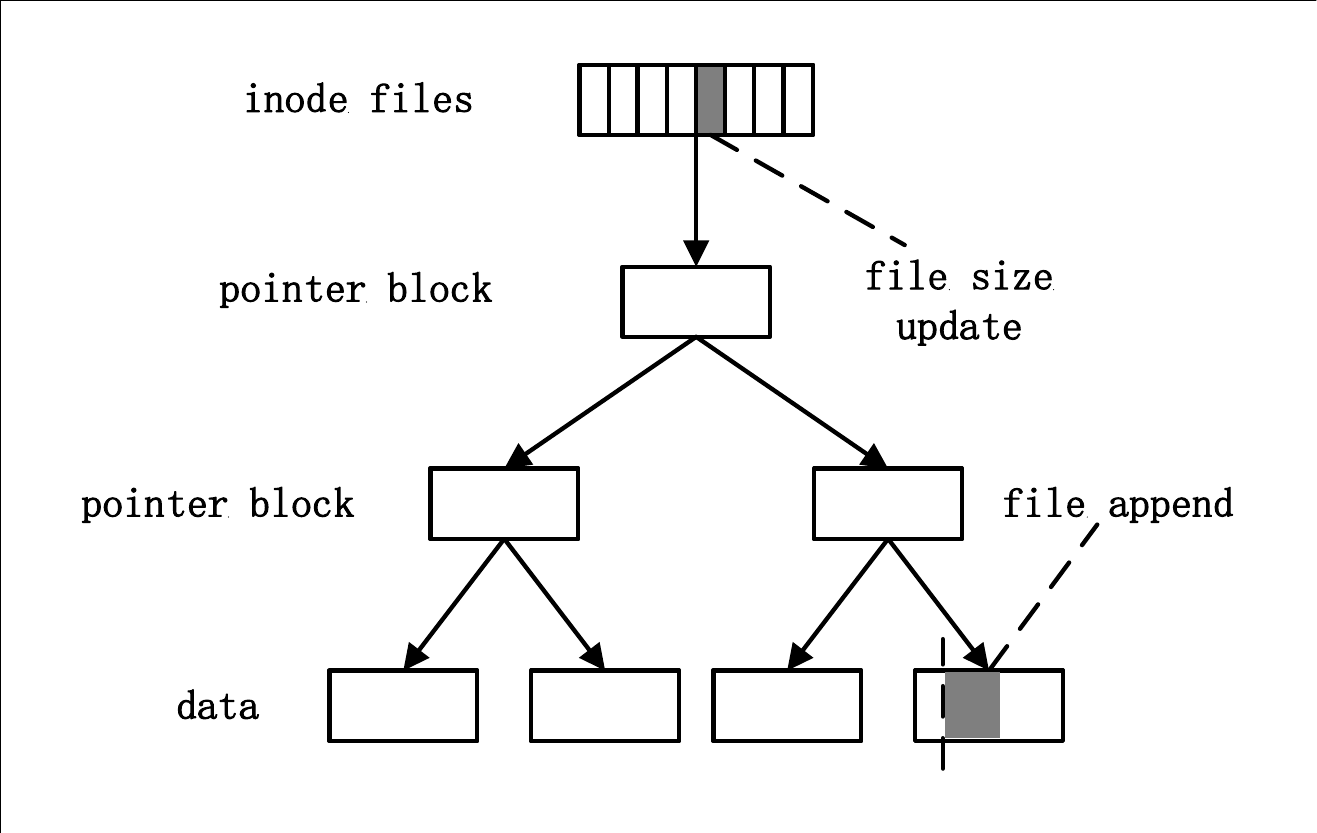}\\
	\parbox[c]{8.3cm}{\footnotesize{Fig.4.~} Shadow paging}
\end{center}

\textbf{Logging Structure.}
Log-structured file systems are designed to exploit higher performance of sequential writes to disk. Random writes are converted into sequential writes in DRAM buffer, and then are synchronized to disk. However, for byte-addressable NVMM, the contiguous free memory region required by logging usually leads to difficulties in memory allocation and garbage collection~\cite{39,FCS2019}. NOVA \cite{39} redesigns the traditional logging structure file system to improve parallelism of data I/O and relax the constrains of contiguous memory allocation. NOVA maintains a log for each updated i-node rather than a uniform contiguous log file. Thus, multiple file updates and recovery can be token in parallel. To mitigate the pressure of contiguous memory allocation and garbage collection operations in log-structured file system, NOVA adopts linked list to store log pages. Besides, to accelerate the access to persistent files, NOVA maintains a directories tree in DRAM. Doshi et al. \cite{41} exploits a backend cache controller to write data to PM  in a asynchronous way. A victim DRAM cache is used to store  cache lines evicted from LLC, and then the persistent updates to NVMM are combined and written to NVMM in a streaming way. CCDS~\cite{16} guarantees data consistency by atomic updating  and maintaining multiple versions of data. After an atomic update to a critical data structure, a new version of the data is created. To ensure data consistent during updating, the most recent version is recorded and can be accessed by all threads.  Upon a power failure, the most recent version is used for recovery while all in-progress updates are removed.

\subsection{Persistent Objects}
As PM-based file system interfaces still rely on complex software I/O stack and can introduce multiple times of data copying, a more attractive way is to store and access application data structures directly in PM, namely \emph{persistent data structures}. 

Persistent data structures have been widely explored object-oriented databases. Berkeley DB~\cite{42} and Stasis~\cite{43} can define persistent data structures explicitly by application programming interfaces (APIs). However, all these systems store the persistent data structures in block-based disks. Recently, there have been a number of persistent object programming frameworks proposed for NVMM, such as NV-Heaps~\cite{21}, NV-Duet~\cite{34} and NVL-C~\cite{74}. Therefore, programmers can definitely allocate NVMM via \emph{pmalloc} and DRAM via \emph{malloc}, and further optimize data placement in hybrid memory systems according to application semantics.

A major challenge of using persistent objects is to guarantee referential integrity of objects, i.e., all references must point to a valid data. Otherwise, a memory leak or a memory error occurs because of dangling pointers or wild pointers. For example, if a pointer in PM refers to an object in DRAM, upon a power failure the pointer in PM becomes a dangling pointer. In a PM system, the memory leak and memory error are usually more destructive since these exceptions may be permanent. The referential integrity may occur in three scenarios: memory allocation, pointer assignment operations and dellocation. In the following, we use NV-Heaps as an example to illustrate those scenarios.

NV-Heaps~\cite{21} is a lightweight and high-performance persistent object systems using the emerging persistent memory. To guarantee data consistency and durability, NV-Heaps provides a set of easy-to-use programming primitives, including persistent objects, specialized pointers, a memory allocator and atomic sections.

For memory allocation, NV-Heaps are also subject to wild pointers as conventional programming models. Once a NV-Heap is not pointed by a valid pointer, its memory space may be permanent unavailable until the NVMM device is reset. To prevent memory leak due to wild pointers, NV-Heaps explores reference counters for garbage collection. A heap is reclaimed immediately once no other objects point to it.

For pointer assignment, four new pointer types may be generated in hybrid memory systems: pointers within an NV-Heap (Intra-Heap NV-to-NV pointers), pointers from an NV-Head to another NV-Heap (Inter-Heap NV-to-NV pointers), pointers from volatile memory to an NV-Heap in NVMM (V-to-NV pointers) and pointers from an NV-Heap in NVMM to volatile memory (NV-to-V pointers). To guarantee referential integrity, NV-to-V pointers and Inter-Heap NV-to-NV should not be assigned. The NV-to-V pointers are unsafe when a program ends. For instance, a pointer \textit{A} in NVMM points to a data structures \textit{B} in DRAM. When the program ends, the memory assigned to \textit{ B} in the DRAM is reclaimed while the persistent pointer \textit{A} remains. An error may occur if the pointer \textit{A} is accessed. The Inter-Heap NV-to-NV pointers are also dangerous. For example, a pointer \textit{P} from NV-Heap \textit{M} points to another NV-Heap \textit{N}. If \textit{N} becomes unavailable, the pointer \textit{P} will point to invalid data. However, the Inter-Heap NV-to-NV pointers are still needed in some cases such as doubly-linked lists. Thus, weak pointers are proposed to implement Inter-Heap NV-to-NV pointers. Weak NV-to-NV pointers act like normal pointers but they do not affect the reference counts. When the reference count becomes zero, all  weak pointers should be atomically released. When an NV-Heap is closed, V-to-NV pointers may lead to a memory leak. To prevent this unsafe closing, NV-Heaps are unmapped only when the program ends.

\subsection{Studies on Intel Optane DCPMMs}

The emergence of Intel Optane DCPMM arises increasing interests in disclosing its performance features and the potential impact on data center applications~\cite{fast20,izraelevitz2019basic,memorysys2019,SC2019,10.1145/3357526.3357568}. Those experimental studies are essential to guide the design of hybrid memory systems and the application programming of DCPMM.  Izraelevitz et al. offer an earliest, scholarly, and comprehensive performance measurements of DCPMM~\cite{izraelevitz2019basic}. They explore its
capabilities as a main memory device, as well as byte-addressable persistent memory exposed to user-space applications.  That report enlightens the research community to understand this non-volatile  memory devices, and to guide future work on hybrid memory systems. They also explore the performance characteristics of Intel Optane DIMMs through both microbenchmarks and
macrobenchmarks, and recommend a set of best practices to maximize the
performance of this device~\cite{fast20}. Weiland et al. explore the performance features of Intel Optane DCPMM and the impact on high-performance
scientific applications in context of performance, efficiency and usability in both Memory and App Direct modes~\cite{SC2019}. A similar work is presented by Patil, et. al~\cite{memorysys2019}. They evaluate the performance characterization of DRAM-NVM hybrid
memory systems for HPC applications using real DCPMM. They find that
the NVMM-only executions are slower than DRAM-only and Memory-mode executions by a
minimum of 2\%, and a maximum of 6X. Peng et al.~\cite{10.1145/3357526.3357568} evaluate the impact of using DCPMM on in-memory
graph processing workloads, and the experimental results suggest that the performance and power efficiency of applications can be optimized by properly distributing data between NVMM and DRAM. 

There have been a few new hybrid memory systems designed from scratch, and also performance optimizations of existing systems using the new Intel Optane DCPMM device. Lersch et al.~\cite{VLDB-index} conducted an extensive study of range indexes on DCPMM. They use an unified programming model for all trees to guarantee fair comparison and developed a benchmarking framework called PiBench. The empirical evaluation has recognized effective techniques, insights,  and caveats to direct the design of future PM-based index structures. Dash~\cite{Dash-VLDB} is a holistic approach to building dynamic and scalable hash tables on DCPMM. The design takes scalability, load factor and recovery into consideration. They develop two popular dynamic hashing schemes, i.e., extendible hashing and linear hashing to demonstrate the efficiency of Dash. Gill et.al.~\cite{10.14778/3389133.3389145}, present the runtime and algorithmic principles of performing large-scale graph analytics on DCPMM and highlight the principles of graph analytics on all large-memory platforms. Mahapatra et al.~\cite{mahapatra2019don} argue that it is inefficient to persist all data structures such as Doubly Linked List, B+Tree and Hashmap in persistent memory. They showcase that alternate partly persistent
implementations can also recreate the data structures along with
the redundant data fields  upon a  system crash. Their solution can significantly improve the performance for a flush-dominated data structure. Ni et al.~\cite{ni-ssrctr-20-01} present performance studies on the interplay of DCPMM hardware and indexing data structures, and propose group flushing and persistent optimized log-structuring techniques for improving the performance of indexing data structure on persistent memories. FlatStore~\cite{FlatStore} is an efficient PM-based key-value store particularly optimized for DCPMM. It decouples the data structure of a KV store into a persistent log structure for efficient storage and a volatile index for fast indexing. Due to the wider availability of DCPMM, more research studies of system design and implementation on real NVMM emerge.

\section{Performance Improvement and Energy Saving}\label{sec:energy}
As NVMMs show much higher access latency and write energy consumption, there have been a lot of studies on performance improvement and energy saving for NVMMs~\cite{CLOCK-DWF,72,73,86,87,park2011power}. These works can be classified into three kinds: reducing the number of NVMM writes, reducing the energy consumption of NVM writes themselves, and DRAM energy consumption reduction.

\subsection{NVMM Write Reduction}
To reduce NVMM writes, a hierarchical architecture is obviously more appropriate since the DRAM cache reduces abundant NVMM writes.  Two major techniques namely page migration and bypassing NVMM writes have been developed for this purpose. 


\textbf{Page Migration.} Page migration \cite{7,8,CLOCK-DWF,93,99} policies choose the pages to be migrated mainly based on the the number of writes and recency of each page. Their main differences are in the conditions in which a page migration is triggered.

PDRAM~\cite{8} migrates PCM pages to DRAM according to the number of writes. In PDRAM, the memory controller maintains a table to record access counts of each PCM page. If the number of writes to a PCM page exceeds a given threshold, a page fault is triggered and then the page is migrated from the PCM page to DRAM.

CLOCK-DWF \cite{CLOCK-DWF} integrates the write history of pages into the CLOCK algorithm. When a page fault occurs, the virtual page is fetched from disk to PCM if it is a read fault. Otherwise, the page is originally allocated in DRAM as the page is likely to be a write-intensive one.

RaPP~\cite{7} migrates pages between DRAM and PCM based on the rank of pages. In RaPP, pages are ranked by the access frequency and recency. Top-ranked pages are migrated from PCM to DRAM. Thus, frequently written pages are placed in DRAM while seldom written pages are placed in PCM. Moreover, RaPP also places mission-critical pages in DRAM to improve application performance. By monitoring the number of writeback operations for each page in LLC, memory controller is able to track the access frequency and recency information of each page. RaPP ranks pages according to the \emph{Multi-Queue} (MQ) algorithm~\cite{10.5555/647055.715773}. A conventional MQ defines $M$ \emph{Least Recently Used} (LRU) queues. Each LRU queue is a queue of page descriptors which include a reference counter and a logical \emph{expiration} time. When a page is accessed at the first time, the page is moved to the tail of the queue 0.
If the reference count of the page reaches $2^{i+1}$, the page is prompted to queue $i+1$. Once a PCM page is moved to the queue 5, it is migrated to DRAM.

\textbf{Buffering NVMM Writes.} In a hybrid memory system, caches are able to reduce a large number of writes to NVMM. A proper cache replacement policy not only improves application performance, but also reduces the energy consumption of NVMM. Previous studies have found that many blocks in cache would not be reused again before they are evicted from the cache. These blocks are called dead blocks and consume precious cache capacity. DASCA \cite{6} proposes a dead block prediction method to reduce the energy consumption of STT-RAM caches. Evicting these dead blocks will reduce the writes to STT-RAM caches and does not affect the cache hit rate. WADE \cite{53} further exploits the asymmetry of energy consumption between NVMM read and NVMM write. As NVMM write operations consume much more energy than NVMM read operations, those frequently-written blocks should be kept in the cache. WADE divides the blocks in cache into two categories: frequently written-back blocks and non-frequently written-back blocks. Non-frequently written-back blocks are replaced to offer more opportunities for keeping other data blocks in the cache.

\subsection{NVMM Energy Consumption Reduction}
Since a NVMM write shows several times higher energy consumption than a NVMM read, there have been many efforts in reducing the energy consumption of NVMM writes. These approaches can be divided into two categories: differential write (only write dirty bits other than the whole line), parallel multiple writes during a single write.

Flip-N-Write~\cite{18} tries to reduce PCM write energy consumption by flipping the bits if the number of bits to be written exceeds half of the total bits in a cache line. During a single write, if more than half of bits in the line are written, each bit is flipped and thus the bit flips are no more than 50\% of total bits. Meanwhile, a tag bit is set to identify whether the bits in a line are flipped.  When the line is read, the tag bit is used to determine whether the bits in the line should be flipped.

Similar to Flip-N-Write, Andrew \cite{17} et. al. advocate fine-grained write.  It only monitors dirty bits rather than all bits in a line. A new term called PCM power token is introduced to indicate the power supply during a single write. Assume each chip is assigned $P_{limit}$ Watts power and each bit-write requires $P_{bit}$ Watts, $\lfloor{P_{limit}/P_{bit}}\rfloor$ bits can be written simultaneously. Within a chip, banks can be written concurrently. During a single write, if a number of write requests located in different banks and the total power consumption doesn't exceed $P_{limit}$, these writes can be executed simultaneously. Thus, fine-grained write not only reduces the NVMM writes, but also improve system performance by achieving higher bank parallelism.

A few studies \cite{72,73} improve energy efficiency of NVMMs by separating the SET and RESET operations. As NVMMs consume more energy and time to write 1 than that of writing 0, both the write latency and energy consumption can be reduced if these writes are performed in a proper manner. Three-stage-write \cite{72} divides a write operation into a comparison stage, a write-zero stage and a write-one stage. In the comparison stage, the Flip-N-Write mechanism is exploited to reduce the number of writes. The zero bits and one bits are written separately in the write-zero stage and the write-one stage, respectively. Because  write-zero operations accounts for a majority of write operations in most workloads, Tetris Write \cite{73} further takes the asymmetry of SET and RESET operations into account, and schedules the costly write-one operations in parallel. The write-zero operations are inserted in the remaining interval of write-one operations under the power constrain.

CompEx \cite{70} proposes a compression expansion encoding mechanism to reduce energy consumption for MLC/TLC NVMMs. To improve the lifetime of MLC/TLC cells, data is compressed first to reduce data redundancy. An expansion code is then applied to the compressed data and written to physical NVMM cells. For a TLC cell with 8 states, the state 0, 1, 6 and 7 are called terminal energy state while the state 2, 3, 4 and 5 are called central energy states. Central energy states consume more time and energy as they need more program and verify iterations. CompEx leverages the expansion code to use only terminal energy state for NVMM cells. This idea is motivated since the terminal energy state  needs less energy and time than the central energy states when programming a MLC/TCL cell. 

Hybrid on-chip caches are also proposed to reduce power consumption of CPUs. RHC \cite{88} constructs a hybrid cache, in which each \textit{way} in SRAM and NVMM can be powered on or off independently. If a row has not been accessed for a long time, the row is powered off while  its tag is still powered on to track the accesses of this row. When the accesses to the tag exceeds a threshold, the row is powered on. To best utilize the high-performance SRAM and the low dynamic-power NVMM, RHC adopts different thresholds for SRAM and NVMM.

\subsection{DRAM Energy Consumption Reduction}
In a memory system with only DRAM, the static energy consumption can accounts for more than half of total energy consumption of memory systems~\cite{54,55,56}. In hybrid memory systems, page migration techniques are widely used to mitigate the energy consumption of DRAM. The inactive pages can be migrated from DRAM to NVMM so that the idle DRAM banks can be powered off. When the page becomes active later, it is migrated to DRAM again. However, if the page migration is not properly performed,  the DRAM ranks may be frequently powered off and reactivate. The extra energy consumption is likely to offset the benefit gained by page migrations.

To reduce energy consumption in hybrid memory systems, RAMZzz~\cite{57} reveals two major roots of high energy cost. One is the sparse  distribution of active pages, another one is that page migrations may be not effective since the transfer among multi-energy states of DRAM introduces additional energy consumption. To solve the former problem, RAMZzz uses multiple queues to collect pages with similar activity into the same DRAM rank, avoiding frequent energy state transfers. The multiple queues have $L$ LRU queues to record the page descriptors. A page descriptor contains the page's ID and access (both read and write) counts in a period of time. To reduce energy overhead of data migration, pages with similar memory access behavior are regrouped together. In this way, pages need to be allocated to new banks. RAMZzz migrates these pages between banks in parallel.

Refree \cite{80} further reduces DRAM energy consumption in hybrid memory systems by avoiding DRAM refresh. When a DRAM row requires to be refreshed, it means the row has not been accessed for a long time. The data in the row is obsolete and it is not likely accessed again in the near further. Refree evicts these rows to PCM rather than refreshing them in DRAM. In Refree, all rows are monitored periodically. The interval of this period is equal to half of the retention time of a DRAM row since its last refresh. Therefore, rows are divided into active rows and non-active rows. The active rows are charged when they are accessed. Non-active rows are evicted to PCM so that DRAM refreshes are eliminated.

\section{Write Endurance Improvement}\label{sec:endurance}

In a hybrid memory system, there are mainly two strategies to overcome the limited write endurance of NVMMs. One is to reduce NVMM writes, and another is wear-leveling which spreads the write traffic evenly among all NVMM cells.

\subsection{Write Reduction}
There have been many write reduction strategies proposed for improving the lifetime of NVMMs, including data migration~\cite{7,8,57}, caching or buffering~\cite{Qureshi:2009}, inner-NVM write reduction~\cite{17,18,36}. 

A lazy write mechanism~\cite{Qureshi:2009} is proposed to reduce the writes in PCM.  In a hierarchical hybrid memory system, a DRAM buffer is used to hide the high-latency PCM accesses. When a page fault occurs, the data is fetched from disk into the DRAM cache directly. The page is not written to PCM until the page is evicted from the DRAM cache. Line-level write can also relieve write operations on NVMMs and thus reduce wear of NVMMs~\cite{Qureshi:2009}. For memory-intensive workloads, the write operations may be concentrated in a few lines. By tracking the cache line in DRAM, only the dirty lines are written back to PCM other than all lines of the page. Memory compression mechanisms~\cite{70,95} are proposed to improve the lifetime of MLC/TLC NVMM.  Data is  compressed first before writing to NVMM cells. Therefore, only a small part of NVMM cells are written. However, the enhancement of endurance is at the expense of a moderate performance degradation.  If a NVMM cell is written with a lower dissipated power, the cell can sustain more writes at the expense of higher write latency. Specifically, When the speed of writing a NVMM cell declines N times, the endurance of the cell can be improved by  $N^1$ to $N^3$ times. Mellow-Writes \cite{82} explores this feature to improve the lifetime of NVMMs. To mitigate the performance degradation, Mellow Writes only adopt slow writes to bank with only one write operation.



\subsection{Wear-Leveling}

Different from write reduction methods, wear-leveling spreads writes among all NVMM pages evenly. Although the total number of write is not reduced, wear-leveling techniques can prevent some pages from being worn out by intensive writes quickly.

For NVMMs, we can record the write counts of each line to guide the wear-leveling policies. However, the external storage overhead can't be ignored. Start-Gap~\cite{58} proposes a fine-grained wear-leveling scheme. The lines of a PCM page are stored in a rotating manner. A rotating value is generated randomly between 0 and 15 to indicate the shifted positions. For a PCM page with 16 lines, the rotating value can range from 0 to 15. When the rotating value is 0, the page is stored in its original address. If the rotating value is 1, line 0 is stored in line 1's physical address, and each line's address is shifted by the rotated value. 

In PDRAM~\cite{8}, wear leveling is triggered by a threshold of write counts. When the write counts of a page exceeds the given threshold, a page swapping interrupt is triggered to migrate the page to DRAM. The swapped PCM page is added to a list in which these pages will be relocated again.

Zombie~\cite{59} offers another direction to achieve weal-leveling, and further extends the overall lifetime of PCM. Other than Start-Gap that distributes writes among PCM cells evenly, Zombie leverages spare blocks in disabled pages to provide more error correction resource for working memory. When a PCM cell is worn out, it becomes unavailable. As memory footprint is organized in pages from the perspective of software, the whole page which contains the failure cell is disabled. However, if some spare cells are provided to replace the failed cells, the page can be used again. These spare cells are called error correction resource. 
When all spare cells are exhausted, the page with failed cells is abandoned finally. Usually, there are about 99\% bits available when a page is disabled. Zombie utilizes the large number of good bits in disabled pages as the spare error correction resource, in which good bits are organized in fine-grained blocks. By pairing the working page with error correction resources, Zombie can extend the lifetime of NVMMs much longer. 

DRM \cite{92} adds an intermediate mapping layer between the virtual address space and the physical NVMM address space. In the intermediate address space, a page may map to a good page in PCM or two compatible PCM pages with faults. The compatible page means a pair of pages with fault bytes, but none of these fault bytes locate in the same place of the two pages. Thus, two compatible pages can be combined to form a new good page. In this way, DRM significantly improves PCM lifetime by 40$\times$.

\section{Practices of Hybrid Memory System Designs}\label{sec:practices}
In this section, we introduce our recent efforts and practices of system designs and optimizations on NVMMs from the perspective of memory architecture, OS-supported hybrid memory management, and NVMM-supported applications, as shown in Figure 5. In the following, we will present those systems briefly. 


\begin{center}
    \includegraphics[trim=7mm 4mm  6mm 6mm, clip, width=\linewidth]{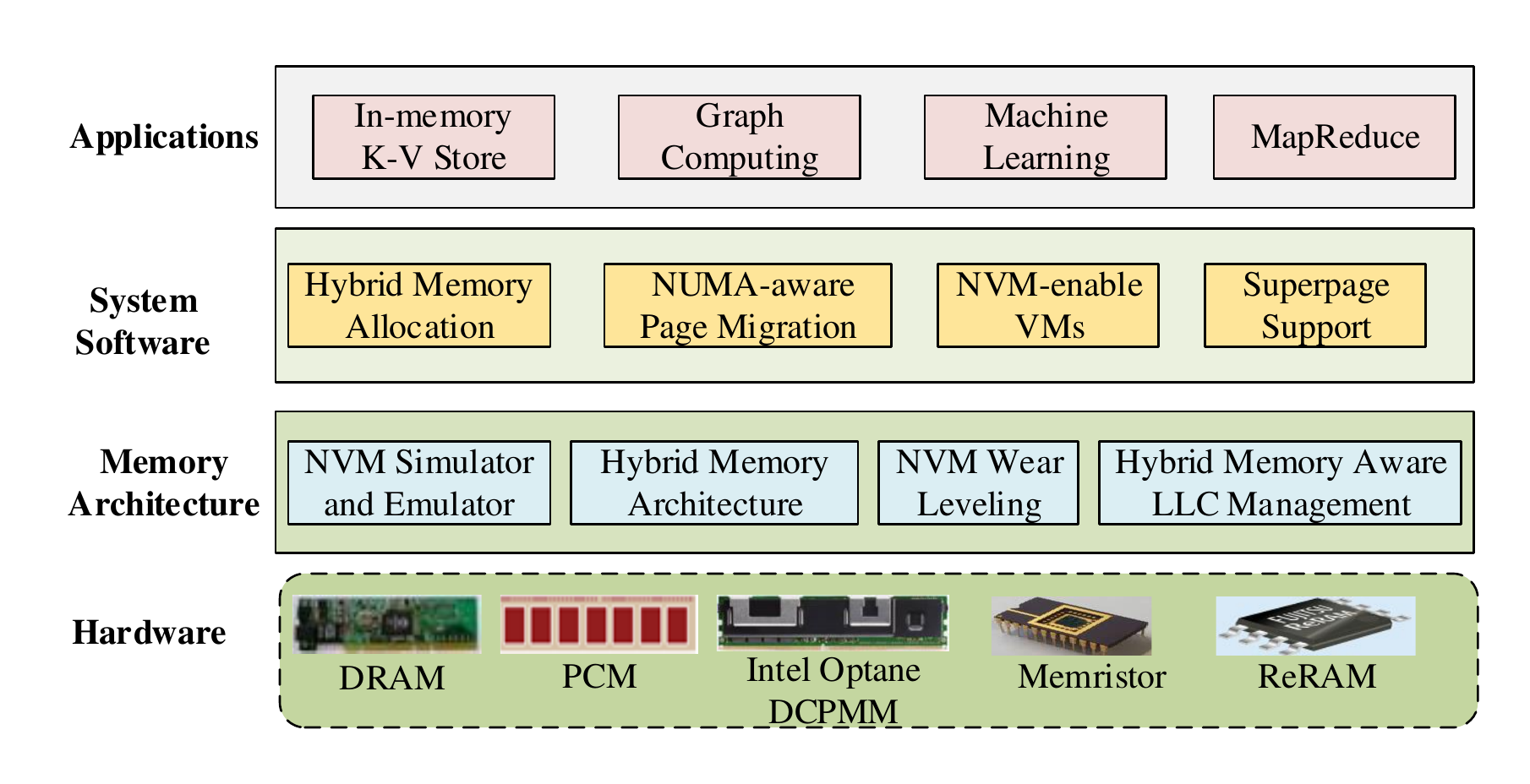}\\
    \vspace{2mm}
    \parbox[c]{8.3cm}{\footnotesize{Fig.5.~} Our practices of system designs on hybrid memories }
\end{center}

\subsection{Memory Architectural Designs}
In this subsection, we present our studies on hybrid memory simulation and emulation, hardware/software cooperative hybrid memory architecture, fine-grained NVM compression and wear leveling, and hybrid memory aware on-chip cache management. 

\subsubsection{Hybrid Memory Architectural Simulation}
A hybrid memory architectural simulation is a prerequisite for studying hybrid memory systems. We integrate zsim~\cite{Sanchez:2013} with NVMain \cite{NVMain} to build a full-system architectural simulator. Zsim is a fast processor simulator for x86-64 multi-core architectures. It is able to model multi-cores, on-chip cache hierarchy, cache coherence protocols such as MESI, on-chip interconnect topology network, and physical memory interfaces. Zsim collects memory trace of processes using Intel Pin toolkit, and then replays the memory trace to characterize the memory access behaviors. NVMain is an architectural Level main memory simulator for NVMMs. It is able to simulate different profiles of memories such as read/write latency, bandwidth, power consumption, and so on. It also support subarray-level memory parallelism and different memory address encoding schemes. Moreover, NVMain can also model hybrid memories such as DRAM and different NVMMs in the memory hierarchy.  Since OS-level memory management is not simulated by zsim, we extend zsim by adding \textit{Translation Lookaside Buffer} (TLB) and memory management modules (such as buddy memory allocator and page tables) to support a full-system simulation. Implementation details are refereed to our open-source software~\cite{HSCC}. Our work provides a fast, and full-system architectural simulation framework to the research community. It can help researchers to understand different NVMM features, design hybrid memory systems, and evaluate the impact of various system designs on application performance in a easy and efficient manner.

\subsubsection{Lightweight NVMM Performance Emulator}

Current simulation-based approaches for studying NVMM technologies are either too slow, or can not run complex workloads such as parallel and distributed applications. We propose HME~\cite{HME}, a lightweight NVMM performance emulator using Non-Uniform Memory Access
(NUMA) architectures. HME exploits hardware performance counters available in commodity Intel CPUs to emulate the performance features of slower NVMMs. To emulate the access latency of NVMMs, HME injects software-generated latency into DRAM accesses
on the remote NUMA nodes periodically. To mimic the NVMM bandwidth, HME utilizes DRAM thermal control interfaces to throttle the amount of memory requests to a DRAM channel in a short period of time. Unlike another NVMM emulator Quartz~\cite{volos2015quartz} that do not emulate the write latency of NVMMs, HME identifies write-through and write-back cache eviction operations and emulates their latencies, respectively. In this way,  HME is able to significantly reduce emulation errors of NVMM access latencies on average compared to Quartz~\cite{volos2015quartz}. Before the advent of real NVMM device--Intel Optane DCPMM, this work can help researchers and programmers to evaluate the impact of NVMM performance characteristics on applications, and guide the system designs and optimizations on hybrid memory systems.

\subsubsection{Hardware/Software Cooperative Caching}
Based on our hybrid memory simulator, we propose a hardware/software cooperative hybrid memory architecture called HSCC. In HSCC, DRAM and NVMM are physically organized in a single memory address space and are all used as main memory. However, the DRAM can be logically used as a cache of NVMM and also managed by OSes.  Figure 6 shows the system architecture of HSCC. We extend page tables and TLB to maintain the NVMM-to-DRAM physical address mappings, and thus manage DRAM/NVMM in the form of a cache/memory hierarchy. In this way, HSCC is able to perform NVMM-to-DRAM address translation as efficient as virtual-to-NVMM address translation. Also, we add an access counter in each TLB entry and page table entry to monitor memory references. Unlike previous approaches monitoring memory accesses in the memory controller or OSes, our design can track all data accesses accurately with trivial storage (SRAM) and performance overhead. We identify frequently accessed (hot) pages through a dynamic threshold adjustment strategy to adapt to different applications, and then the hot pages in NVMM are migrated to DRAM cache for higher performance and energy efficiency.  Moreover, we develop an utility-based DRAM cache filling scheme to balance the efficiency of DRAM cache and DRAM utilization. As the software-managed DRAM pages are able to map to any NVMM pages, the DRAM is actually used as a fully associative cache. This approach can significantly improve the utilization of DRAM cache, and also offers opportunities to reconfigure the hybrid memory architecture according to dynamic memory access behaviors of applications. As CPUs can bypass the DRAM cache to directly access cold data in NVMM, the DRAM can be used either as main memory in the flat-addressable hybrid memory architecture, or as a data filter cache of NVMM in the hierarchical hybrid memory architecture. As a result, HSCC can significantly improve system performance by up to 9.6X and reduce energy consumption by about 34.3\% compared to a state-of-the-art work~\cite{Qureshi:2009}. Our work offers the first architectural solution to achieve reconfigurable hybrid memory systems that can dynamically change the management of DRAM/NVMM between horizontal and hierarchical memory architectures. 

\begin{center}
    \includegraphics[width=\linewidth]{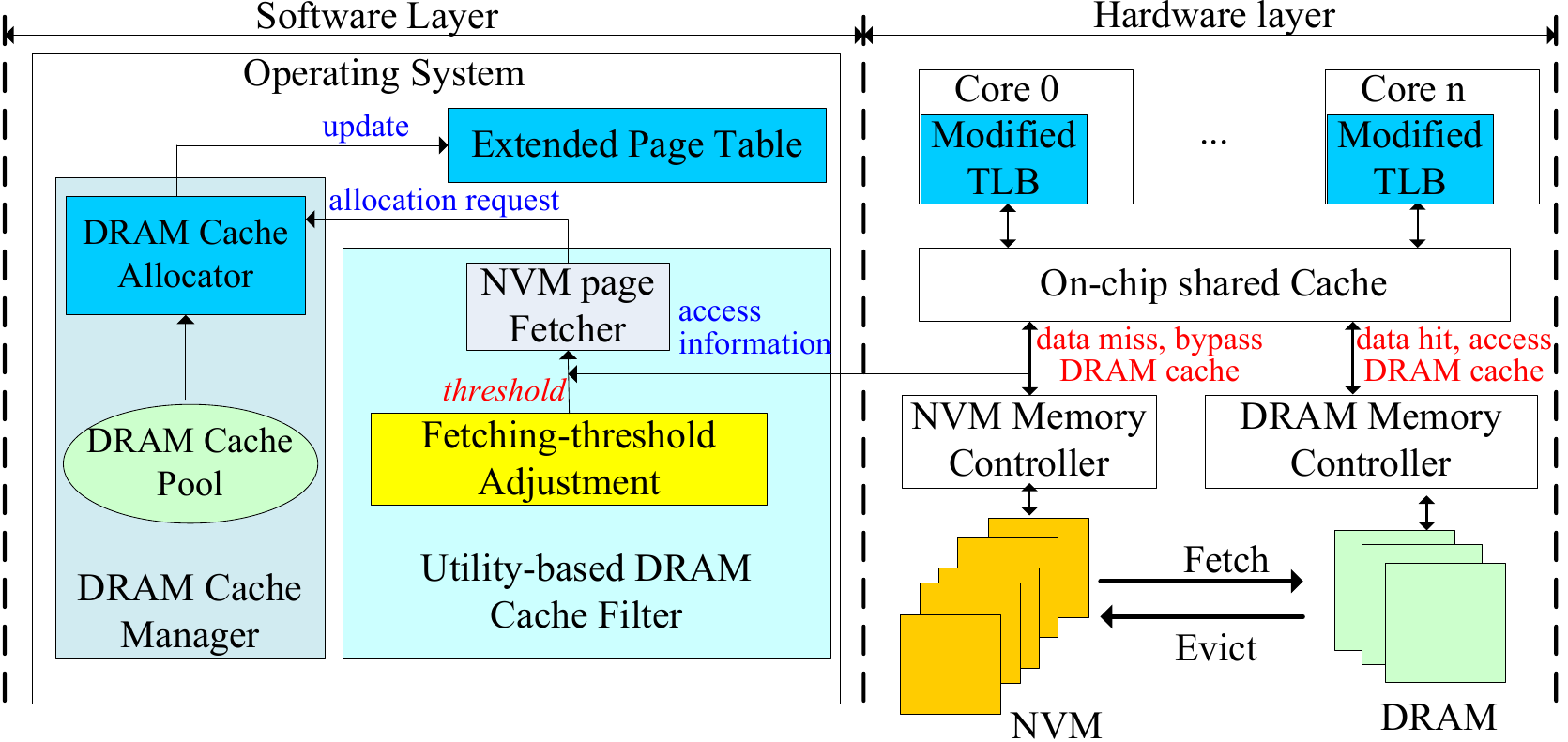}\\
    \vspace{2mm}
    \parbox[c]{8.3cm}{\footnotesize{Fig.6.~} System architecture of HSCC~\cite{Liu:2017}}
\end{center}


We further propose the following techniques on HSCC to improve the cache performance and improve the wear-leveling mechanisms. 

As the cache miss penalty for a NVMM block is several times higher than a DRAM block, the cache hit rate is not the only one performance metric that should be improved in a flat-addressable hybrid memory architecture. To best utilize the expensive LLC, we propose a new metric, i.e., \emph{Average Memory Access Time} (AMAT), to assess the overall performance of hybrid memory systems.  We take the asymmetrical cache miss penalty of DRAM blocks and NVMM blocks into account, and propose a LLC miss penalty aware replacement algorithm called MALRU~\cite{MALRU-DATE,MALRU-TCAD} to improve the AMAT in hybrid memory systems. MALRU partitions the LLC into a reserved region and a normal replacement region dynamically. MALRU preferentially replaces dead and cold DRAM blocks in the LLC so that NVMM blocks and hot DRAM blocks are kept in the reserved region. In this way, MALRU achieve application performance improvement by up to 22.8\% compared with the LRU algorithm. This work showcases how the hybrid memory system can effect the architectural design of on-chip cache.


To improve the write endurance of NVMM, we propose a new NVMM architecture to support space-oblivious data compression and wear-leveling. As memory blocks of many applications usually contain a large amount of zero bytes and frequent values, we propose \textit{Zero Deduplication} and \textit{Frequent Value Compression} mechanisms (called ZD-FVC) to reduce bit-writes on NVMM. ZD-FVC can be integrated into the NVMM module and implemented entirely by hardware, without any intervention of Operating Systems. We implement ZD-FVC in Gem5 and NVMain simulators, and evaluate it with several programs from SPEC CPU2006. Experimental results shows that ZD-FVC is much better than several state-of-the-art approaches. Particularly, DZ-FVC can improve data compression ratio by 1.5X compared to \textit{Frequent Value Compression}. Compared with \textit{Data Comparison Write}, ZD-FVC is able to reduce bit-writes on NVMM by 30\%, and significantly improve the lifetime of NVMM by 5.8X on average. Correspondingly, ZD-FVC also reduces NVMM write latency by 43\% and energy consumption by 21\% on average. Our design provides a fine-grained data compression and wear-leveling solution for NVMMs in simple and efficient manner. It is complementary to other wear-leveling schemes to further improve NVMM lifetime.

\subsection{System Software for Hybrid Memories}

In this subsection, we present our practices of hybrid memory systems in the software layer, including object-level hybrid memory allocation and migration, NUMA-aware page migration, superpage supporting, and NVMM virtualization mechanisms.

\subsubsection{Object Migration in Hybrid Memory Systems}
\begin{figure*}[tbp]
    \centering
    \includegraphics[scale=0.55]{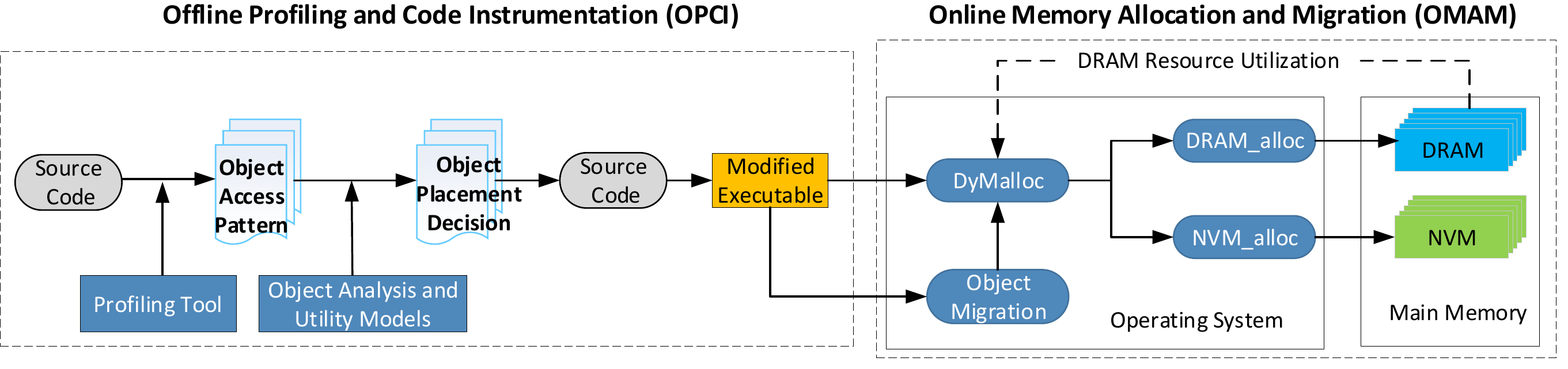}
    \parbox[c]{8.3cm}{\footnotesize{Fig.7.~} An overview of OAM framework~\cite{TC2020}}
    \vspace{-3ex}
\end{figure*}

Page migration techniques have been widely exploited to improve system performance and energy efficiency in hybrid memory systems. However, previous page migration schemes all rely on costly online page access monitoring schemes in the OS layer to track page access recency or frequency. Moreover, data migration at the page granularity often results in non-trivial performance overhead because of additional memory bandwidth consumption and cache/TLB consistency guarantee mechanism. 

To mitigate the performance overhead of data migration in hybrid memory systems, we propose more lightweight object-oriented memory allocation and migration mechanisms called OAM~\cite{TC2020}. The framework of OAM is shown in Figure 7. Unlike previous studies~\cite{Hassan:2015,Exploiting} that only profiles memory access behaviors in a global view for static object placement, we further analyze object access patterns in fine-grained time slots. OAM leverages a compile framework LLVM to profile application memory access patterns at the object granularity, and then divides the execution of applications into different phases.  OAM exploits a performance/energy integrated model to guide the initial memory allocation and runtime object migration in different execution phases, without intrusive modifications of hardware and OSes for online page access monitoring. We develop new memory allocation and migration APIs by extending the \textit{Glibc} library and Linux kernel. Based on these APIs, programmers are able to  allocate DRAM or NVMM to different objects explicitly, and then migrate the objects whose access patterns are dynamically changed between DRAM and NVMM.  We develop a static code instrumentation tool to automatically modify legacy applications' source codes, without re-engineering the applications by programmers. Compared to the state-of-the-art page migration approaches CLOCK-DWF~\cite{CLOCK-DWF} and 2PP~\cite{Exploiting}, experimental results show that OAM can significantly reduce data migration cost by 83\% and 69\% respectively, meanwhile achieve about 22\% and 10\% application performance improvement. Previous persistent memory management schemes often rely on memory access profiling to guide static data placement, and page migration (costly) techniques to adapt to dynamic memory access patterns at runtime. OAM provide a more lightweight hybrid memory management scheme which provides fine-grained object-level memory allocation and migration.   

\subsubsection{NUMA-aware Hybrid Memory Management}
In {Non-Uniform Memory Access} (NUMA) architectures, application-observed  memory access latencies in different NUMA nodes are usually asymmetrical. Because NVMM is several times slower than DRAM, hybrid memory systems can further enlarge the performance gap among different NUMA nodes. Traditional  memory management mechanisms for NUMA systems is no longer effective in  hybrid memory system and may even degrade application performance. For example, The \textit{automatic NUMA balancing} (ANB) policy always migrates application data in a remote NUMA node to a NUMA node in which the application threads or processes are running. However, since the access performance of remote DRAM may be even higher than that of local NVMM, ANB may falsely move application data to a slower place. To address this problems, we propose \textit{HiNUMA}~\cite{ICCD2019}, a new NUMA abstraction for hybrid memory management. When application data is first placed in the hybrid memory system,  \textit{HiNUMA} places application data on both NVMM and DRAM to balance memory bandwidth utilization and total access latency for bandwidth-sensitive applications and latency-sensitive applications, respectively. The initial data placement are based on NUMA topology and hybrid memory access performance.  For runtime hybrid memory management, we propose a new NUMA balancing policy named \textit{HANB} for page migrations. \textit{HANB} is able to reduce the total cost of hybrid memory accesses by taking both data access frequency and memory bandwidth utilization into account. We implement HiNUMA in Linux kernel, without any
modifications of hardware and applications. Compare with traditional memory management policies in NUMA architectures and other state-of-the-art works,  \textit{HiNUMA} can significantly improve application performance by efficiently utilizing hybrid memories. The lessons learned from this work is also applicable to hybrid memory systems equipped with real Intel Optane DCPMM device.

\subsubsection{Supporting Superpages in Hybrid Memory Systems}
With a rapid growth of application footprint and the corresponding memory capacity, virtual-to-physical address translation has become a new performance bottleneck for hybrid memory systems. superpages have been widely exploited to mitigate address translation overhead in big-memory systems. However, the side effect of using superpages is that they often impede lightweight memory management such as page migration, which is widely exploited in hybrid memory systems to improve system performance and energy efficiency. Unfortunately, it is challenging to have both world of superpages and  lightweight page migration.

To address this problem, we propose a novel hybrid memory management systems called \textit{Rainbow}~\cite{TACO2019} to bridge the fundamental conflict between superpages and lightweight page migration. As shown in Figure 8, Rainbow manages NVMM at the granularity of superpages (2 MB), and manages DRAM as a cache to store hot data blocks in superpages at the granularity of base pages (4 KB). To speed up address translations, Rainbow employs the existing hardware feature of split TLBs to support superpages and normal pages. 
\begin{center}
	\includegraphics[width=0.9\linewidth]{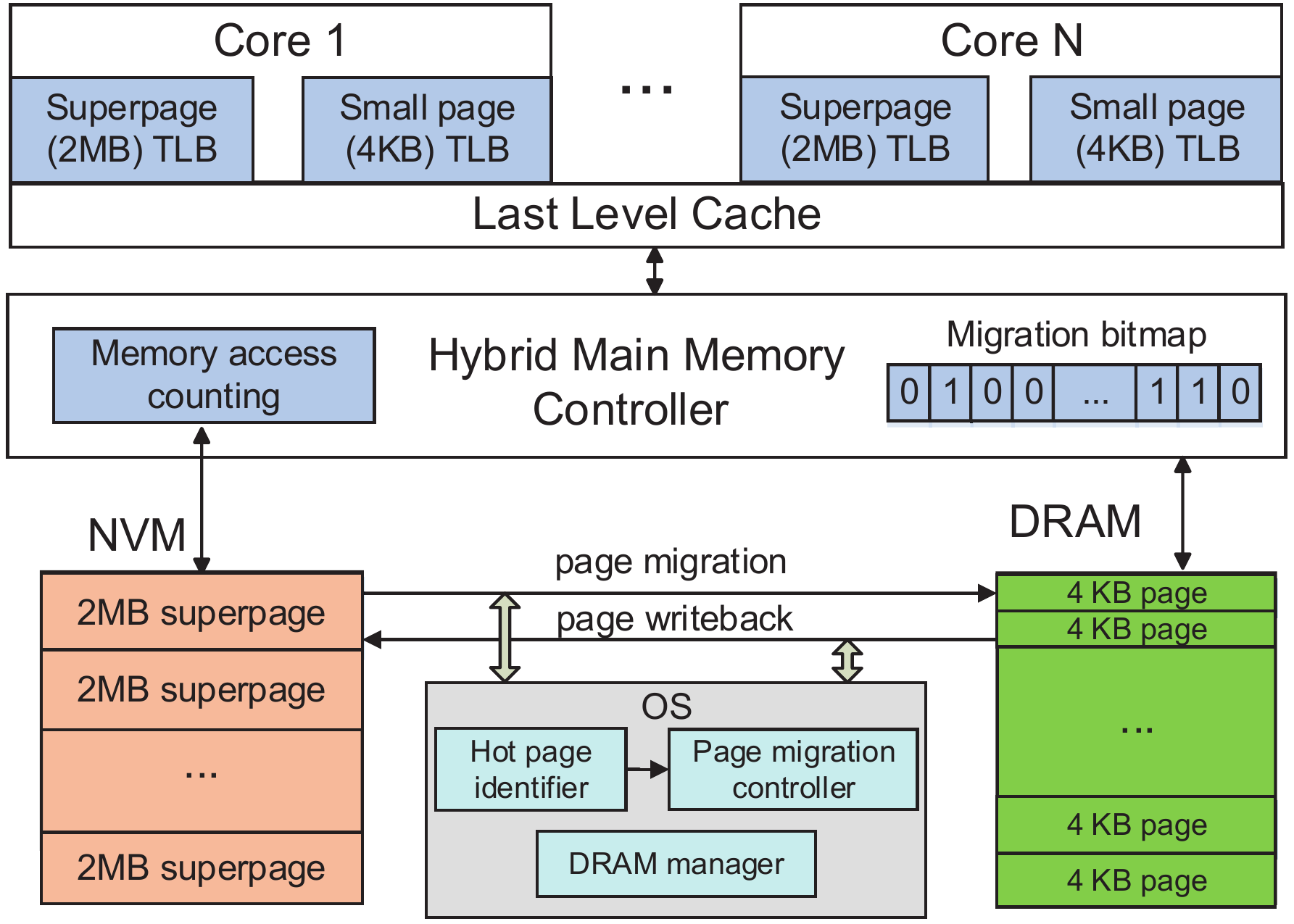}\\
	\vspace{2mm}
	\parbox[c]{8.3cm}{\footnotesize{Fig.8.~} The architecture of Rainbow~\cite{TACO2019}}
\end{center}

We propose a two-stage page access monitor mechanism to identify hot base pages within superpages. In the first stage, Rainbow records the access counts of all superpages to identify $top$-$N$ hot superpages. In the second stage, we logically split those hot superpages into base pages (4 KB) and further monitor them to recognize hot base pages. This schemes significantly diminish the SRAM storage overhead for page access counters and runtime performance overhead due to sorting the hot base pages. With a new NVMM-to-DRAM address remapping mechanism, Rainbow is able to migrate hot base pages to DRAM while still guaranteeing the integrity of superpage TLB. The split superpage TLBs and base page TLBs are consulted in parallel. Our address remapping mechanism logically uses superpage TLBs as a cache of the base page TLBs. Because the hit rate of superpage TLB is often very high, Rainbow is able to significantly accelerate base page address translation. To further improve TLB hit rate, we also extend Rainbow to support multiple page sizes and migrate contiguous hot base page together~\cite{ACCESS2020}. Compared with a state-of-the-art hybrid memory system without superpage support, Rainbow can significantly improve application performance by at most 2.9X by having the benefit of both using superpages and lightweight page migration. 

This work provides a hardware/software cooperative design to bridge the fundamental conflict between superpages and lightweight page migration techniques. This may be a promising solution to mitigate the ever-increasing virtual-to-physical address translation overhead in large-capacity hybrid memory systems.  

\subsubsection{NVMM Management in Virtual Machines}

NVMMs are expected to be more popular in cloud and data center environments. However, there have been few work on using NVMMs for \textit{virtual machines} (VMs). We propose \textit{HMvisor}~\cite{SCIS2019}, a hypervisor/VM cooperative hybrid memory management system to utilize DRAM and NVMM efficiently. As shown in Figure 9, HMvisor exploits a pseudo-NUMA mechanism to support hybrid memory allocation in VMs.  Since virtual NUMA nodes in a VM can be mapped to different physical NUMA nodes, HMvisor can map different memory regions to a single VM and thus expose memory heterogeneity to VMs. 

\begin{center}
	\includegraphics[width=\linewidth]{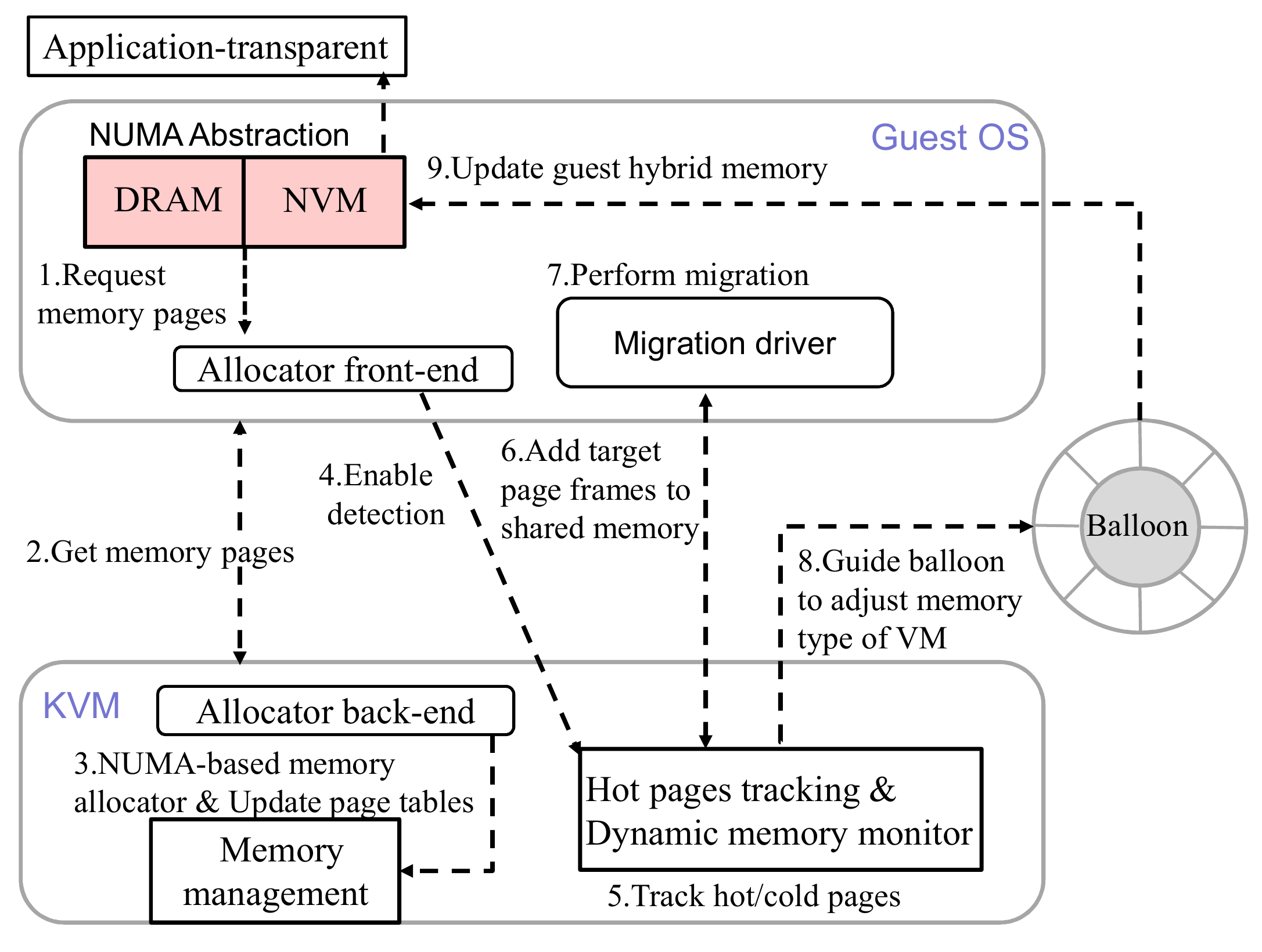}\\
	\vspace{2mm}
	\parbox[c]{8.3cm}{\footnotesize{Fig.9.~} System overview of HMvisor~\cite{SCIS2019} }
\end{center}

To support lightweight page migration in VMs, HMvisor monitors page access counts and classify hot pages and cold pages in the hypervisor, and then the VM periodically collects the information of hot pages through a inter-domain communicate mechanism. We implement a loadable driver in the VM to perform process-level page migrations between DRAM and NVMM. Since HMvisor performs page migrations by the VM itself, HMvisor does not need to suspend the VM for page migrations. HMvisor also advocates a hybrid memory resource trading policy to dynamically adjust the size of NVMM and DRAM in a VM. In this way, HMvisor can meet different memory requirement (capacity or performance) of diversifying applications while keeping the total monetary cost of the VM unchanged.  

The prototype of HMvisor is implemented in the QEMU/KVM platform. Our evaluation shows that HMvisor is able to reduce NVMM write traffic by 50\% at the expense of only 5\% performance overhead. Furthermore, the dynamic memory adjustment policy can significantly reduce major page faults in a VM when it suffers high memory pressure, and thus can even improve application performance by 30 times. 

This is an early system work to manage hybrid memory in a virtualization environment. The proposed schemes are completely implemented by software, and thus also applicable to hybrid memory systems comprising of the new Intel Optane DCPMM device.

\subsection{NVMM-supported Applications}

Since hybrid memory systems can provide very large capacity of main memory, they have been widely explored for big data applications such as in-memory key-value (K-V) stores and graph computing. In this subsection, we present our practices of NVMM-supported system optimizations for those applications.

In-memory K-V stores with large-capacity memory can cache more hot data in main memory, and thus deliver higher performance to applications. However, there are several challenges to directly deploy traditional K-V stores such as \textit{memcached} in hybrid memory systems. For example, how to identify hot K-V objects efficiently? How to redesign a NVMM-friendly K-V indexes to reduce NVMM writes? How to redesign the cache replacement algorithm to balance object access  frequency and recency in hybrid memory systems. How to address the slab calcification problem~\cite{hu2015lama} to best utilize DRAM resource in hybrid memory systems. 

To address the above problems, we propose \textit{HMCached}~\cite{HMcached}, an extension of K-V cache (memcached) for hybrid DRAM/NVMM systems. Figure 10 shows the system architecture of HMCached. HMCached tracks K-V object accesses and records the access counts in each K-V pairs' metadata structure, so that HMCached can easily identify frequently-accessed (hot) objects in NVMM and migrates them to DRAM. In this way, we logically use DRAM as an exclusive cache of NVMM to avoid more costly NVMM accesses. Moreover, we redesign an NVMM-friendly K-V data structure by splitting the hash-based K-V indexes to further reduce NVMM accesses. We put the frequently-updated metadata (e.g., reference counts, timestamp, and access counts) of K-V objects in DRAM, and the remaining portion (e.g. keys and values) in NVMM. We exploit a multi-queue algorithm to take both object access frequency and recency into accounts for DRAM cache replacement. Moreover, we set up an utility-based performance model to evaluate the benefit of slab class reassignment. Our dynamic slab reallocation policy is able to address the \textit{slab calcification} problem effectively, and significantly improve application performance when the data access pattern changes. Compared to the vanilla memcached, HMCached can significantly reduce NVMM accesses by 70\% and achieves about 50\% performance improvement. Moreover, HMCached is able to reduce 75\% DRAM cost while the performance degradation is less than 10\%. 
 
To the best of our knowledge, we are the first to explore object-level data management for K-V stores in hybrid memory systems. We implement HMCached based on Memcached and open the source codes. We find that later works such as flatstore~\cite{FlatStore} have a similar idea to decouple the data structure of KV stores.
 
 \begin{center}
    \includegraphics[width=\linewidth]{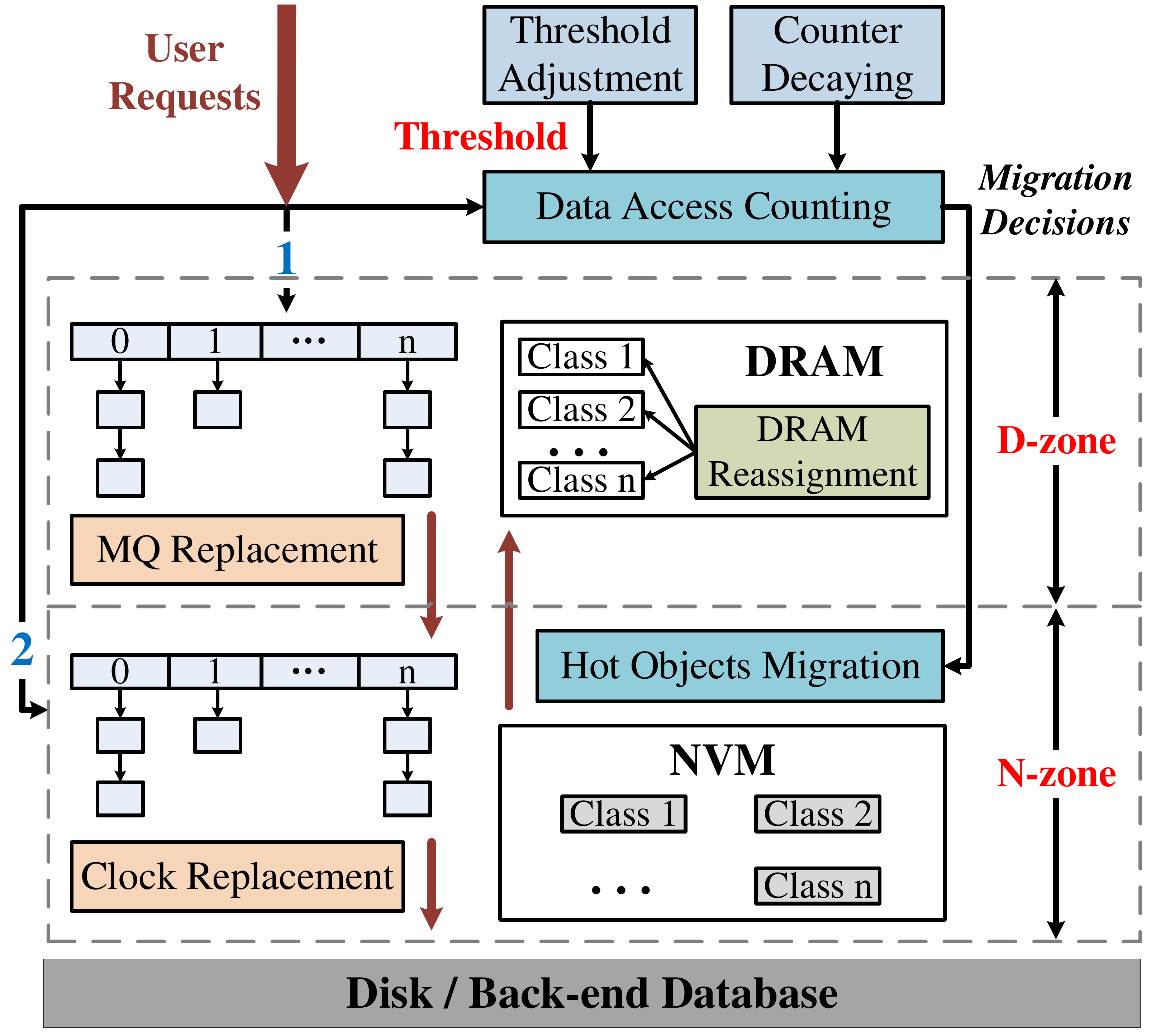}\\
    \vspace{2mm}
    \parbox[c]{8.3cm}{\footnotesize{Fig.10.~} Architecture of HMCached~\cite{HMcached}}
\end{center}
 
Today we have witnessed a number of in-memory graph processing systems in which application performance are highly bound to the capacity of main memory. High-density and low-cost NVMM technologies are essential to mitigate I/O cost for graph processing. As shown in Figure 11, hybrid memory system can significantly improve application performance compared to a SSD-based storage system. Figure 12 shows the application performance gap between a hybrid memory system and a DRAM-only system. Although the gap is acceptable, it indicates that there are still opportunities to further exploit the advantages of NVMM and DRAM for in-memory graph processing systems. 

\begin{center}
	\includegraphics[width=0.67\columnwidth]{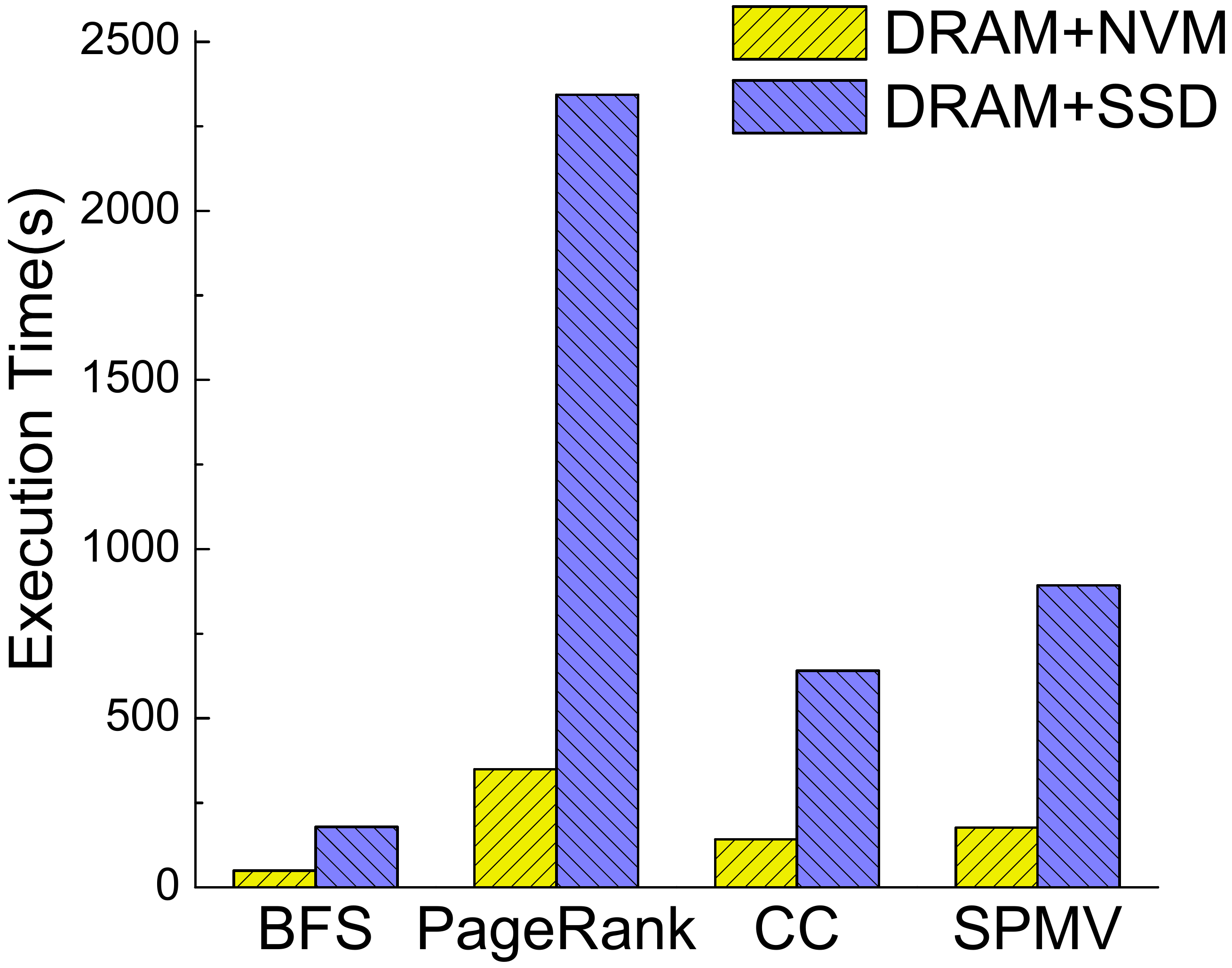}   \\
	\vspace{2mm}
	\parbox[c]{8.3cm}{\footnotesize{Fig.11.~} Performance difference between ``DRAM+NVM'' and ``DRAM+Disk'' systems}
\end{center}

\begin{center}
	\includegraphics[width=0.67\columnwidth]{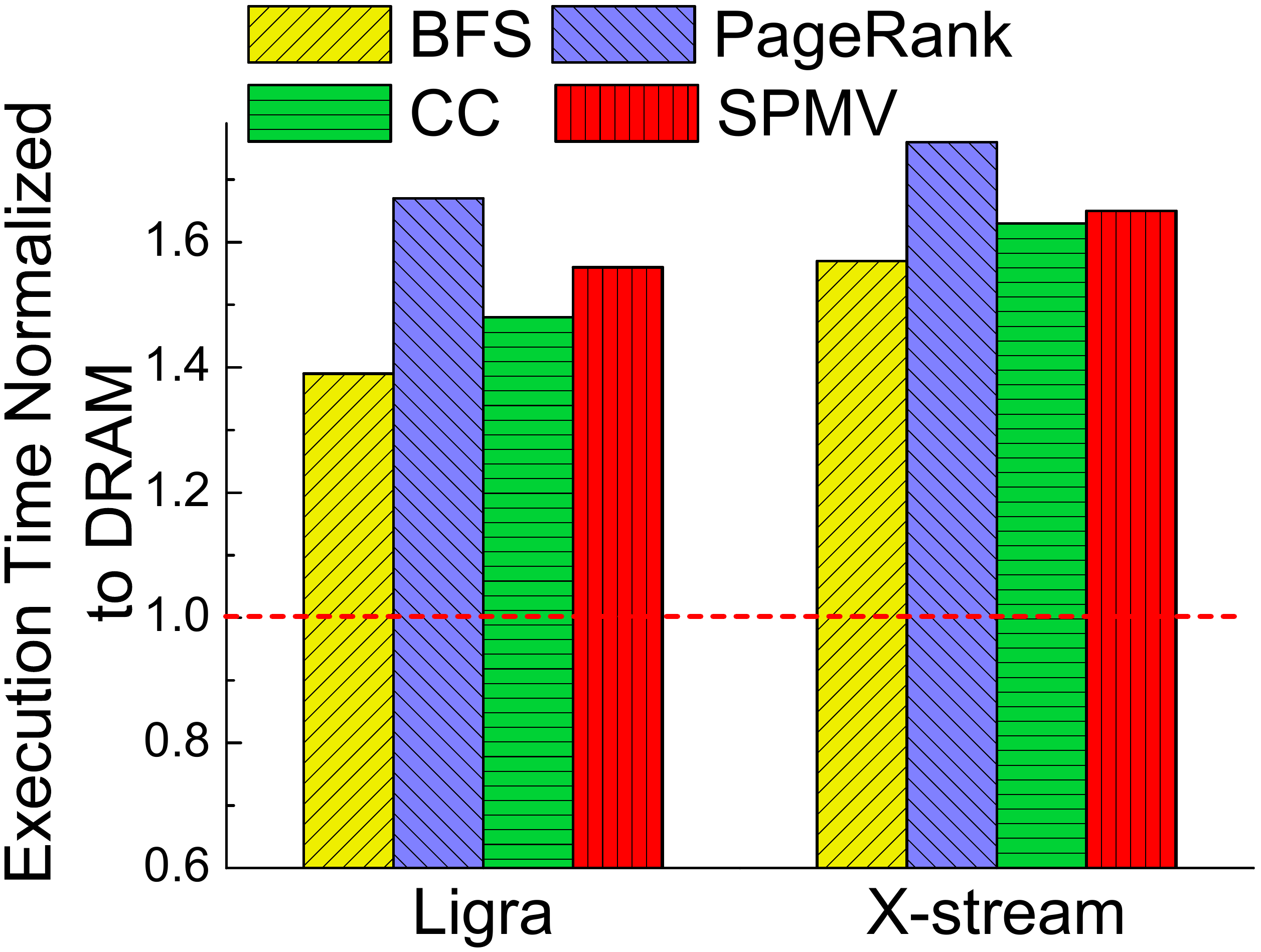} \\
	\vspace{2mm}
	\parbox[c]{8.3cm}{\footnotesize{Fig.12.~} Performance difference between  a hybrid memory system and a DRAM-only system }
\end{center}

We propose \textit{NGraph}~\cite{NGraph}, a new graph processing framework particularly designed to better utilize hybrid memories.  We develop hybrid memory aware data placement policies based on access patterns of different graph data to mitigate random and frequent accesses to NVMM. Generally, graph structure data accounts for a majority of total graph data. NGraph partitions graph data according to destination vertices and employs a task decomposition mechanism to avoid data contention between multiple processors. Moreover, NGraph adopts a work stealing mechanism to minimize the maximum time of parallel graph data processing on multicore systems. We implement NGraph based on a graph processing framework Ligra.  NGraph can improve application performance by up to 48\%  compared with the state-of-the-art work Ligra. The lessons learned from this work can be exploited to further improve the performance of large-scale graph analytics in a graph processing platform equipped with real PM device.

\section{Future Research Directions}\label{sec:vision}
The advent of NVMM technologies has aroused many interesting research topics in the area of material, microelectronics, computer architecture, system software, programming model, and big data applications. As real NVMM device such as Intel Optane DCPMM~\cite{Intel-Optane-DIMM} has been increasingly applied to data center environments, NVMMs may change the storage landscape of data centers. Our experiences and practices have demonstrated some preliminary and interesting studies on those dimensions. In the following, we share our vision of future research directions of NVMMs, and analyze the research challenges and new opportunities. Figure 13 illustrates the future trends of NVMM technologies in different dimensions.

 \begin{center}
    \includegraphics[width=\linewidth]{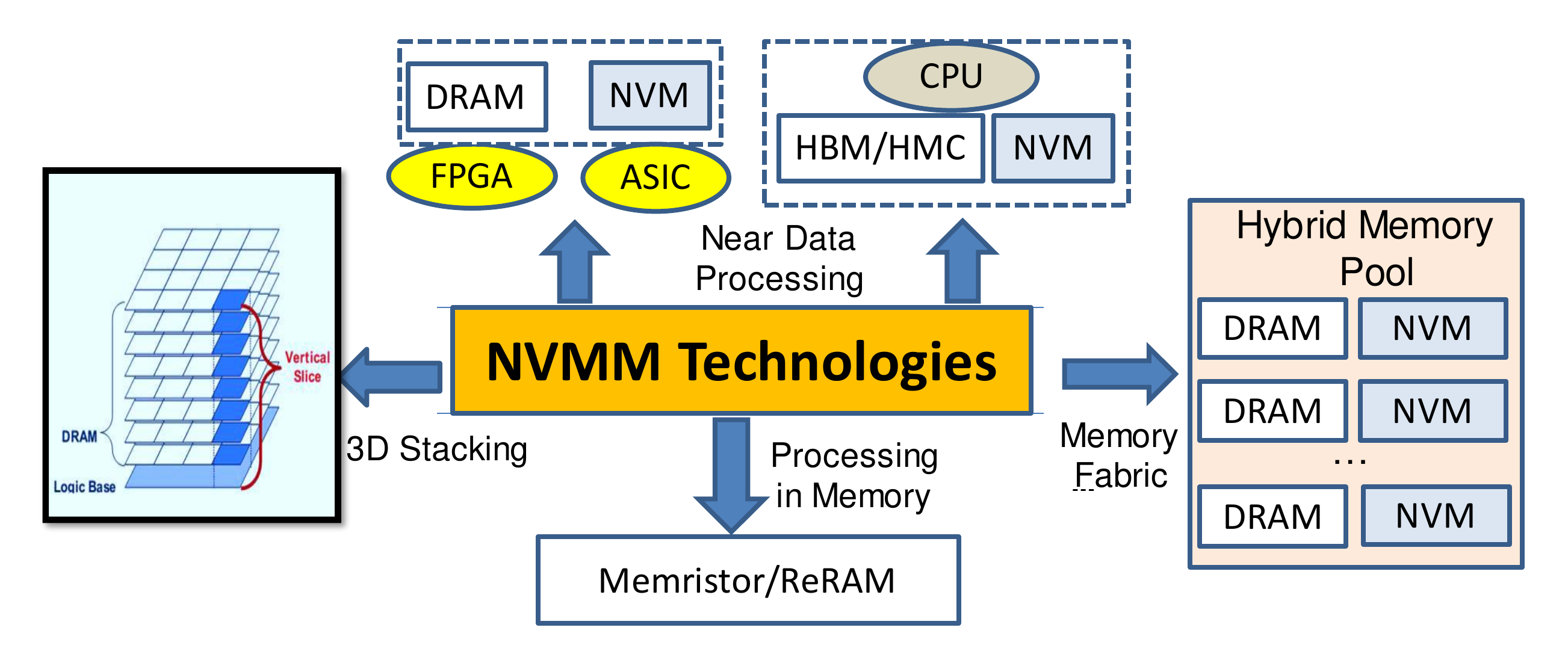}\\
    \vspace{2mm}
    \parbox[c]{8.3cm}{\footnotesize{Fig.13.~} Future directions of NVMM technologies }
\end{center}

\textit{First, the development of of 3D stacked NVMM technologies is still continuing.} NVMMs are expected to provide higher integration density for cost reduction.  Currently, the high-end NVDIMMs is still too expensive for enterprise applications. The key challenge for NVMMs to compete with traditional DRAM and NAND flash is the storage density or the cost per byte. There are mainly two monolithic 3D integration mechanism for NVMM technologies~\cite{7495087}. One is to stack the horizontal cross-point array layer by layer, such as Intel/Micron 3D X-point~\cite{Intel-Optane-DIMM}. Another is the vertical 3D stacked structure that is referred to ReRAM technology. However, the 3D integrate technologies are not full-blown. There still remains many challenges such as fabrication cost,  pillar
electrode resistance, and sneak path problems.

\textit{Second,  NVMMs are increasingly used in distributed shared memory systems.}  As the density of NVMMs is continuously increasing, the main memory capacity can approach hundreds of Terabytes in a single server. To improve the utilization of large-capacity NVMMs, it is essential to share them among multiple servers via \textit{remote direct memory access} (RDMA) techniques. A typical approach of using NVMMs is to aggregate all shareable memories from multiple servers in a hybrid shared memory resource pool, such as Hotpot~\cite{Orion,Octopus,hotpot}. All memory resources are shared in a global memory space. There have been a few preliminary studies on using NVMMs in datacenter and cloud environments~\cite{Orion,hotpot,FCS2020}. A new trend of using PM is to manage it as disaggregated memory, like traditional disaggregated storage systems. This model is different from the previous shared PM systems in which the PM DIMMs are distributed in multiple servers and shared by user-level applications. These computation-memory tightly-coupled architectures has several drawbacks in terms of manageability, scalability, and resource utilization. In contrast, the disaggregated PM systems equip a large amount of PM in a few memory nodes with less computations, can are connected by computation nodes via high-speed fabric. This computation/memory disaggregated architecture can mitigate the above challenges in data center environments more easily.  However, there still remain a lot of challenges. For example, the persistent feature of NVMMs also should be guaranteed in distributed environments. Traditional PM management instructions such as \textit{clflush} and \textit{mfence} can only guarantee data persistence in a single server, but can not guarantee data persisting to a remote server over RDMA networks. For each RDMA operation, once the data arrives to the \textit{network interface card} (NIC) in the remote server, it issues an acknowledgment to the data sender. As there are data buffers in NICs, the data is not  stored to the remote NVMM immediately. If a power failure occurs at this time, data persistence is not guaranteed. Thus, it is essential to redesign the RDMA protocol to support flushing primitives. Moreover, the computation nodes should support remote page swapping which should be transparent to user-level applications. To support this mechanism, the traditional virtual memory management policies should be redesigned. On the other hand, since PM shows memory-like performance and are byte-addressable, new designs on memory scheduling and management are required to adapt to disaggregated PM.

\textit{Third, NVMM-based computation/memory integrated computer architectures are arising.} For example, the use of emerging NVMMs in \textit{processing-in-memory} (PIM)~\cite{Pinatubo,ReRAM-PIM} and \textit{near data processing} (NDP)~\cite{HotOS2017,Biscuit} architectures are arising. PIM and NDP have emerged as new computing paradigms in recent years. NDP refers to the integration of a processor with memory on a single chip so that the computation can access the data in memory as closer as possible. NDP is able to significantly reduce the cost of data movement. There are mainly two approaches to this goal. One is to integrate small computation logics such as (FPGA/ASIC) into memory chips so that data can be pre-processed before it is finally fetched to CPUs.  Another approach is to integrate memory units (HBM/HMC) into computation (CPUs/GPGPUs/FPGAs) . This model is commonly used by many processor architectures such as Intel Xeon Phi Knights Landing series, NVIDIA tesla V100, and Google Tensor Processing Unit (TPU). PIM refers to processing data entirely in computer memory. It offers high bandwidth, massive parallelism, and high energy efficiency by performing computations in main memory. PIM using NVMMs (such as ReRAM) usually can compute the bitwise logic of two or more memory rows in parallel, and support one-step multi-row operations. This paradigm is particular efficient for matrix-vector multiplication in an analog computing manner, and can achieve an extremely large degree of performance speedup and energy saving. As a result, PIM is widely explored in accelerating machine learning algorithms such as \textit{convolutional neural networks} (CNN) and \textit{deep neural network} (DNN). Although there are growing interests in using NVMM technologies in PIM architectures~\cite{DNN-NVM,8715178,Pinatubo,ReRAM-PIM}, current works are mainly based on electrical simulations, and none of them are available for mid-scale prototyping.

\textit{Fourth, beyond the traditional applications, some novel applications using NVMMs are emerging.} Although NVMM technologies have been preliminarily adopted in a lot of big data applications, such as K-V store, graph computing, and machine learning. However, most of those programming frameworks/models and runtime systems are designed for disk devices and DRAM based main memory, they are not effective and efficient in hybrid memory systems. For example, buffering and lazy-write mechanisms are widely utilized in those systems to hide the high latency of I/O operations. However, those mechanisms may be not needed in hybrid memory systems and may even hurt application performance. These big data processing platforms such as Hadoop/Spark/GraphChi/Tensorflow should be redesigned to adapt to the features of NVMM technologies. Beyond those traditional applications, some novel applications based on NVMMs are emerging. For example, there have been a few proposals to use NVMMs as hardware security primitives such as physical unclonable functions (PUFs) by exploiting the intrinsic variations of NVMM's switching processes~\cite{6998001}. PUFs are typically used in applications with high security requirements, for example, cryptography. Recently, a number of logic circuits based on NVMM technologies have been proposed and prototyped~\cite{6571792,6177067,10.1145/1596543.1596548}. For example, the ReRAM technology is proposed to use as reconfigurable switch for ReRAM-based FPGAs~\cite{6177067}. Moreover, the STT-RAM technology is proposed to design non-volatile cache or registers~\cite{10.1145/2463585.2463592}.

\section{Conclusions}\label{sec:conclusion}
Emerging NVMM technologies have many good features relative to traditional DRAM technologies. They have a potential to fundamentally change the landscape of memory systems and even add new functionalities and features to the computer systems. There are vast opportunities
to rethink the designs of todays' computer systems to achieve orders
of magnitude improvement in system performance and energy consumption. This paper presents a comprehensive survey of the state-of-the-art works and our practices from the perspective of memory architecture, OS-level memory management, and application optimizations. We also share our vision of future research directions about NVMM technologies. By taking advantage of the unique features of NVMMs, there are enormous opportunities to innovate the future's computing paradigm and develop a lot of diverse novel applications of NVMMs.

%

\section*{Acknowledgments}
This work is supported jointly by National Natural Science Foundation of China (NSFC) under grants No. 61672251, 61732010, 61825202, 61929103.

\bibliographystyle{JCST}
\bibliography{bibitem}

\label{last-page}
\end{multicols}
\label{last-page}
\end{document}